\documentclass[a4paper,11pt]{article}
\pdfoutput=1 
\usepackage{jheppub}
\usepackage[T1]{fontenc} 
\usepackage{mathtools}
\usepackage{setspace}
\usepackage{slashed}
\usepackage{amsfonts}
\usepackage{amssymb}
\usepackage{graphicx}
\usepackage{caption}
\usepackage{subcaption}
\usepackage{array}
\usepackage{booktabs}
\usepackage{multirow}
\usepackage[normalem]{ulem}

\title{\boldmath $\frac{|V_{ub}|}{|V_{cb}|}$ and Quest for New Physics\\
}

\author[a,b]{Anshika Bansal}
\author[a]{Namit Mahajan}
\author[a,b]{Dayanand Mishra}

\affiliation[a]{Physical Research Laboratory,\\Ahmedabad, India.}
\affiliation[b]{Indian Institute of Technology, \\Gandhinagar, India.}

\emailAdd{anshika@prl.res.in}
\emailAdd{nmahajan@prl.res.in}
\emailAdd{dayanand@prl.res.in}

\arxivnumber{2112.00363}

\abstract{Charged current semi-leptonic decays of $B$-meson are important for a precise determination of the CKM elements.  
	 The CKM elements $|V_{ub}|$ and $|V_{cb}|$ show a discrepancy between the exclusive
	 and inclusive determinations. These determinations are however masked with hadronic and other uncertainties, and thus can't
	 be unambiguously taken as implying new physics. 
	In the present study, we propose a new observable: the ratio of these two CKM elements,  $R_V \equiv \frac{|V_{ub}|}{|V_{cb}|}$,
	which is found to receive negligible corrections due to hadronic as well as QED effects.
	Interestingly, $R_V$ as constructed from exclusive determinations of $|V_{ub}|$ and $|V_{cb}|$ agrees quite well with that
	constructed from the inclusive determinations of these CKM elements.
	Hence, we propose that $R_V$ is a better and cleaner observable, and can serve as an excellent tool for the test of the Standard Model.
	We also provide an example of its probing power.}

\begin{document} 
\maketitle
\flushbottom

\section{Introduction}
	\label{sec1}
	The Cabibbo-Kobayashi-Maskawa (CKM) matrix is a unitary matrix which governs the mixing of different flavours of quarks in the 
	Standard Model (SM). The four independent parameters of the CKM matrix are the fundamental parameters of the SM, and hence a 
	precise determination of these elements is important to validate the SM, as well as to probe physics beyond it. The study of 
	charged current and neutral current induced processes thus plays a crucial role in achieving the same. Charged current processes 
	are present at tree level while flavour changing neutral current processes are present only at the loop level in the SM. \\
	 At the quark level, a $B$-meson decays to a $D$ or $\pi$-meson via an exchange of virtual $W$ boson which can further decay 
	 to $\ell$-$\nu_\ell$ pair. The strength of these semileptonic $B$-meson decays  is governed by the CKM elements 
	 $\left|V_{cb}\right|$ and $\left|V_{ub}\right|$, respectively. Their amplitude can be factorised into the leptonic and the hadronic parts which helps in factoring out the hadronic
	 uncertainties coming due to our lack of understanding of strong interactions. Hence they provide an excellent ground to 
	 measure the CKM elements $|V_{cb}|$ and $|V_{ub}|$ 
	 \cite{Bourrely:2008za,Bailey:2008wp,Monahan:2018lzv,Jaiswal:2017rve,Grinstein:2017nlq,Na:2015kha,Biswas:2021qyq,Bernlochner:2017jka,Wang:2017jow}. Alternatively, these CKM elements can also be extracted from inclusive decays,
	 $B\to X_{c,u} \ell \nu_{\ell}$ \cite{Lange:2005yw,Gambino:2007rp,Leibovich:1999xf,Bauer:2002sh,Gambino:2013rza,
	 Alberti:2014yda,Khodjamirian:2011ub,Boyd:1995cf,Ricciardi:2019zph,Alberti:2016fba}. Other exclusive modes can also be considered but will not be the focus 
	 of the discussion below.\\ 
	 On the experimental front, $|V_{cb}|$ and $|V_{ub}|$ show discrepancy of $\sim 3\sigma$ and $\sim 3.5\sigma$, 
	 respectively between the inclusive and the exclusive determinations. These discrepancies are popularly known as the 
	 $|V_{cb}|$ and $|V_{ub}|$ puzzles (or 'exclusive' vs 'inclusive' puzzles) 
	 \cite{HFLAV:2019otj,Zyla:2020zbs,Aoki:2021kgd,Gambino:2020jvv}. Whether the origin of such a discrepancy in 
	 $|V_{cb}|$ and $|V_{ub}|$ is a hint of new physics, or simply a consequence of the underestimation of the theoretical 
	 and/or experimental uncertainties is still an open question \cite{Ricciardi:2021shl,Faller:2011nj,
	 Crivellin:2014zpa,Bigi:2015uba,Colangelo:2016ymy,Bigi:2017njr,Gambino:2019sif,Akeroyd:2003zr,Chen:2008se,
	 Crivellin:2012ye,Crivellin:2009sd,Buras:2013ooa}. The origin of theoretical uncertainties essentially lies in computing non-perturbative 
	 quantities entering the respective decay modes reliably, while at the same time applying suitable kinematical cuts.\\
	Though theoretical uncertainties get lowered by using more precise form factors calculated using Light Cone Sum Rules (LCSRs) and 
	lattice QCD, their complete removal seems a nearly impossible task with our current understanding and capabilities of handling strong interactions. 
	Hence, one looks for observables where these hadronic uncertainties can be removed or significantly minimised. In view of this, 
	many lepton flavour universality (LFU) ratios have been proposed in literature like $R_{K^{(*)}}$, $R_{D^{(*)}}$
	\cite{Hiller:2003js,Chen:2006nua,Bhattacharya:2014wla,Bordone:2016gaq,Bobeth:2007dw,Hiller:2014yaa,Fajfer:2012vx,Kamenik:2008tj}. Here,  $R_{K^{(*)}}$ is defined by the 
	ratio of the branching ratio of $B\to K^{(*)} \mu \mu$ to the branching ratio of $B\to K^{(*)} e e$ while $R_{D^{(*)}}$ is 
	defined by the ratio of branching ratio of $B\to D^{(*)} \tau \nu$ to the branching ratio of $B\to D^{(*)} \mu \nu$. 
	Though these ratios are less sensitive to hadronic uncertainties by construction, but what about the soft photon QED corrections? The experimental analysis partially includes the effect of soft photons using the PHOTOS Monte-Carlo generator 
	\cite{Barberio:1993qi,Golonka:2005pn,Davidson:2010ew}. However, the contributions
	like the emission of photons depending on the structure of hadrons, the interference between the initial and final 
	state emissions and virtual corrections are not included in PHOTOS. To have an understanding of the complete dynamics 
	inclusion of these contribution becomes important. It has been found that on inclusion of these contributions, 
	the LFU ratios are not 
	free from soft photon QED corrections\cite{Misiak:2010zz,Bernlochner:2010yd,Becirevic:2012jf,deBoer:2018ipi,Cali:2019nwp,
	Isidori:2020acz,Mishra:2020orb}, 
	particularly when photon energy and/or angular cuts have to be explicitly specified. This leads us to the quest for	
	observables which should be less sensitive to hadronic uncertainties as well as the QED corrections due to soft photons.\\
	Experimental analysis has been performed for the ratio of the CKM elements, $\frac{|V_{ub}|}{|V_{cb}|}$ using two different
	modes: ($1$) the baryonic modes ($\Lambda_{b}^{o}\to p \mu^{-}\bar{\nu_{\mu}}$ and $\Lambda_{b}^{o}\to \Lambda_{c}^{+} \mu^{-}\bar{\nu_{\mu}}$)
	leading to $|V_{ub}|/|V_{cb}|=0.083\pm 0.004$ \cite{LHCb:2015eia,Detmold:2015aaa}; and 
	($2$) the mesonic modes ($B_{s}^{o}\to K^{-} \mu^{+}\bar{\nu}_{\mu}$ and $B_{s}^{o}\to D_{s}^{-} \mu^{+}\bar{\nu_{\mu}}$)
	giving $|V_{ub}|/|V_{cb}|=0.095\pm 0.008$ ($0.061\pm 0.004$) for high $q^{2}$ (or low $q^{2}$) \cite{LHCb:2020ist}. 
Also, it is interesting to note that, using the PDG values \cite{Zyla:2020zbs}, the ratio $\frac{|V_{ub}|}{|V_{cb}|}$ formed for exclusive determinations of $|V_{ub}|$ and $|V_{cb}|$ is in fantastic 
	agreement with that for inclusive determinations. 
	\begin{eqnarray}
	\frac{|V_{ub}|}{|V_{cb}|}\Big|^{\text{high $q^2$}}_{\text{excl}}=0.094\pm0.005 \hspace{1cm} \frac{|V_{ub}|}{|V_{cb}|}\Big|^{\text{high $q^2$}}_{\text{incl}}=0.101\pm 0.007 
	\label{eqn1s}
	\end{eqnarray}
	 Intrigued, and motivated by this, we consider the ratio $\frac{|V_{ub}|}{|V_{cb}|}\equiv R_V$ in the present study.
	We show that this ratio gets negligible corrections from soft photons and explicitly check that it is minimally
	affected by the choice of the form factors adopted for $B\to D $ and $B\to \pi$ transitions.\footnote{It should be borne in mind that the experimental extractions above differ in the low- and high- $q^2$ regions due to differences in the
	employed form factors. This is due to the fact that currently the form factors derived with different approaches 
	are known with better accuracy in different $q^2$ regions. Thus, it is important to choose the $q^2$ range judiciously to ensure
	that the observable is least affected by the choice employed. It is in this sense we mean independent of form factor choice.}
	We therefore suggest the use of $R_V$ in phenomenological studies as it is a cleaner observable compared to the usual LFU ratios, 
	and can have better potential in probing new physics.\\ 
	The rest of the paper is organised as follows: in Section-\ref{sec2}, we discuss the decay width of the non-radiative process $B\rightarrow P \mu \nu_\mu$ decay where $P=\pi$ or $D$-meson. In Section-\ref{sec3}, we discuss the impact of soft photon QED corrections to $B\rightarrow P \mu \nu_{\mu}$ decay due to real and virtual photon emission at $\mathcal{O}(\alpha)$. In this section, we also discuss the photon inclusive case and the phase space structure for this radiative decay. In Section- \ref{sec4}, we discuss our results for the effect of soft photon correction on the observable $R_V$ and sensitivity to the choice of form factors. Finally, we conclude in Section-\ref{sec5} and discuss the implications of the use of $R_V$ for beyond the standard model searches.

\section{Non-radiative $B\to P\ell\nu_{\ell}$ ($P=D,\pi$)}
	\label{sec2}
	Consider the process $B(p_{B},m_{B})\to P(p_{P},m_{P}) \ell(p_{\ell},m_{\ell}) \nu_{\ell}(p_{\nu},0)$ where, P is a pseudo-scalar meson (D or $\pi$). If the final state massless particles are left unobserved, then the second order differential decay rate for this process can be described fully by two independent Lorentz invariant variables
	\begin{eqnarray}
	y=\frac{2 p_{B}.p_{l}}{m_{B}^{2}}, \text{  and  }\hspace{0.3cm}  z=\frac{2 p_{B}.p_{P}}{m_{B}^{2}}.
	\label{eqn1}
	\end{eqnarray}
	One can also choose the Mandelstam variables $q^{2}=(p_{B}-p_{P})^{2}\equiv m_{B}^{2}+m_{P}^{2}-2 p_{B}.p_{P}$ and $s_{B\ell}=(p_{B}-p_{\ell})^{2}\equiv m_{B}^{2}+m_{\ell}^{2}-2 p_{B}.p_{\ell}$ instead of $y$ and $z$. The amplitude for $B\rightarrow P \ell \nu_\ell $ can be factorised into the hadronic and the leptonic contributions as
	\begin{eqnarray}
	\mathcal{M}_{0}(B\to P\ell\nu_{\ell})=\frac{G_{F}}{\sqrt{2}}V_{qb}\mathcal{H}_{\mu}(p_{P},p_{B})\mathcal{L}^{\mu} .
	\end{eqnarray}
	Here $|V_{qb}|$ $(q={c,u})$ is the CKM matrix element and $G_{F}$ is the Fermi constant. $\mathcal{L}^{\mu}(=u_{\ell}\gamma^{\mu}(1-\gamma^{5})v_{\nu_{\ell}})$ and $\mathcal{H}_\mu$ are the leptonic and the hadronic matrix elements, respectively. $\mathcal{H}_{\mu}$ can be parametrized in terms of two $q^{2}$ dependent form factors, $f_{+}^{P}$ and $f_{0}^{P}$ as
	\begin{eqnarray}
	\mathcal{H}_{\mu}(p_{P},p_{B})=(p_{B}+p_{P}) f_{+}^{P} +(p_{B} - p_{P}) f_{-}^{P}
	\end{eqnarray} 
	where $f_{-}^{P}=\frac{m_{B}^{2}-m_{P}^{2}}{q^{2}}(f_{0}^{P} - f_{+}^{P})$. These form factors are computed via different methods like
	LCSRs or lattice QCD, and often 
        employing different parametrization. However, the choice of form factors does not play a significant role for the determination of $\frac{|V_{ub}|}{|V_{cb}|}$ (see Sec- \ref{sec4}). \\
For the present study, we have chosen the model independent parametrization for $B\to D \ell \nu_{\ell}$ and z-expansion parametrization for $B\to \pi \ell \nu_{\ell}$. The explicit form of these form factors in these parametrizations are given in Appendix-\ref{appb}. It is to be noted that these form factors for both the processes are valid for the entire $q^2$ range.

The total decay width for the non-radiative process $B\rightarrow P \ell \nu_\ell$ reads
	\begin{eqnarray}
		\Gamma^{0}&=& \frac{m_{B}}{256 \pi^{3}}\int dz \int dy \left|\mathcal{M}_0\right|_{B\to P \ell \nu_{\ell}}^{2},
		\label{nondefde}
	\end{eqnarray}
		\begin{eqnarray}
		\text{where}\hspace{1cm}
	\left|\mathcal{M}_0\right|_{B\to P \ell \nu_{\ell}}^{2}= \frac{G_{F}^{2}}{2}\left|V_{qb}\right|^{2}\left((f_{0}^{P})^{2} c_{1} + (f_{+}^{P})^{2} c_{2} + f_{0}^{P}f_{+}^{P} c_{3}\right),
	\end{eqnarray}
	with the coefficients $c_i$ (where $i=1,2,3$) given by
	\begin{eqnarray}
	c_{1}&=& -\frac{4(m_{B}^{2}-m_{P}^{2})^{2}m_{l}^{2}\Big((z-1)m_{B}^{2}+m_{l}^{2}-m_{P}^{2}\Big)}{(m_{P}^{2}-(z-1)m_{B}^{2})^{2}},\nonumber\\
		c_{2}&=& -\frac{4m_{B}^{2}}{(m_{P}^{2}-(z-1)m_{B}^{2})^{2}}\Bigg[-(z-1)m_{B}^{4}\Big(m_{l}^{2}(4y(z-2)+3z^{2}-8z+8)+4 m_{P}^{2}(2 y^{2}+2 y (z-2)\nonumber \\ 
	&-& 3z+3)\Big)+m_{B}^2(m_{P}^{2}m_{l}^{2}(4y(z-2)+3z^{2}-4z+4)+(z-2)^{2}m_{l}^{4}+4 m_{P}^{4}(y^{2}+y(z-2) -3 z + 3)) \nonumber\\ 
	&+& 4(y-1)(z-1)^{2}m_{B}^{6} (y+z-1) - 4 m_{P}^{2} m_{l}^{2}+ 4 m_{P}^{2}\Bigg], \text{  and}\nonumber\\
	c_{3}&=&\frac{8m_{B}^{2}(m_{B}^{2}-m_{P}^{2})m_{l}^{2}\left((z-1)m_{B}^{2}(2y+z-2)-(z-2)m_{l}^{2}-m_{P}^{2}(2y +z-2)\right)}{(m_{P}^{2}-(z-1)m_{B}^{2})^{2}}.
	\end{eqnarray}
	 We now discuss the effect of soft photon emission on this decay width.
    	\section{Soft photon QED Corrections to $B \rightarrow P \ell \nu_\ell$}
	\label{sec3}
 Here, we have two possibilities for charge assignment  of the particles: first is where $P$ is charged and $B$ is neutral, and second where $B$ is charged and $P$ is neutral. The computation for the soft photon corrections in the two cases is analogous with minute difference in the selection of the kinematic variables. We present the case of $B^- \rightarrow P^0 \ell^- \bar{\nu}_\ell$ here and point out the necessary differences for the case of $B^0 \rightarrow P^+ \ell^- \bar{\nu}_\ell$ wherever required.
	\subsection{Correction due to Real photon emission}
	\label{subsec1}
\begin{figure}[h]
	 	\begin{subfigure}{.5\textwidth}
	 		\centering
	 		\includegraphics[width=0.3\linewidth]{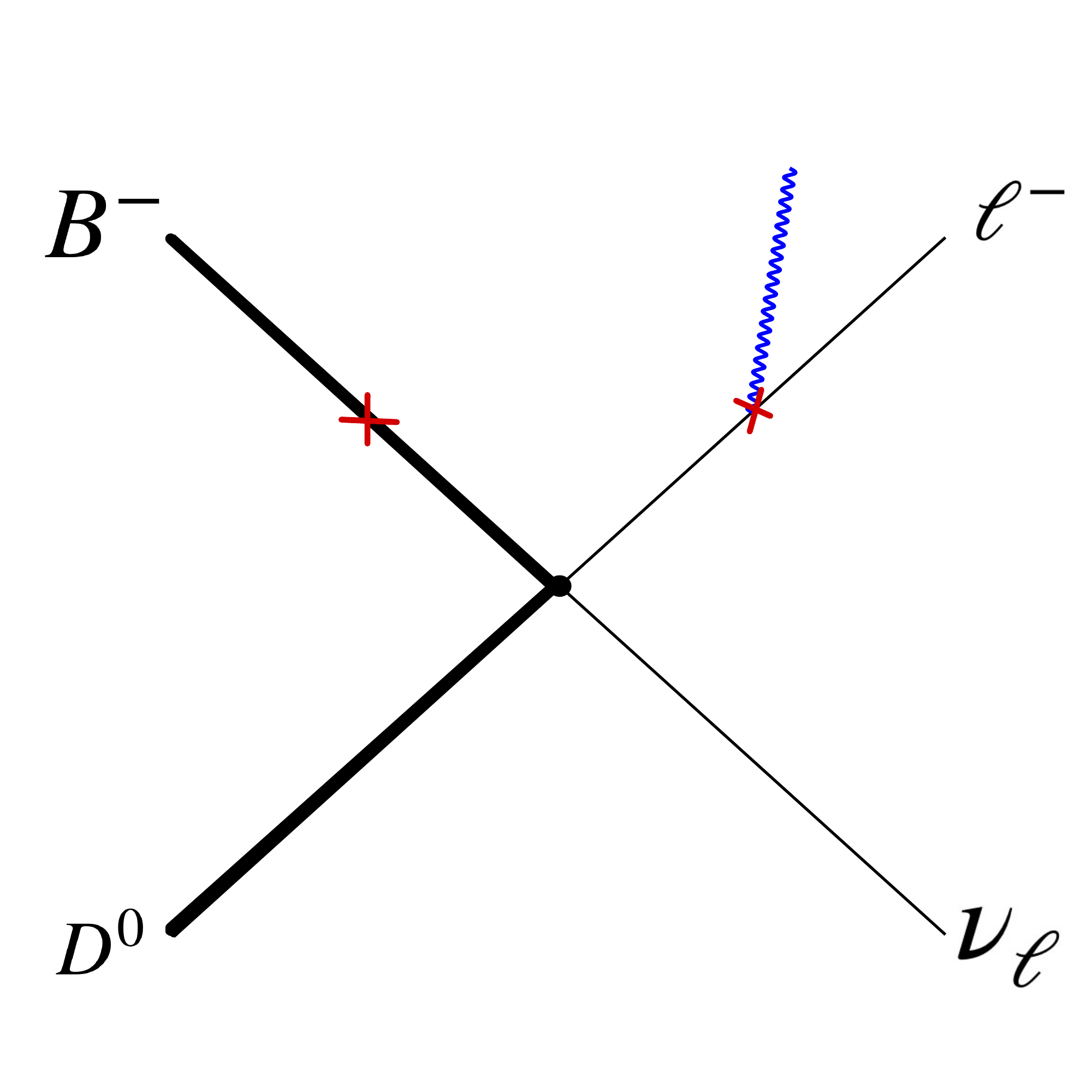}
	 		\caption{}
	 	\end{subfigure}%
	 	\begin{subfigure}{.5\textwidth}
	 		\centering
	 		\includegraphics[width=0.3\linewidth]{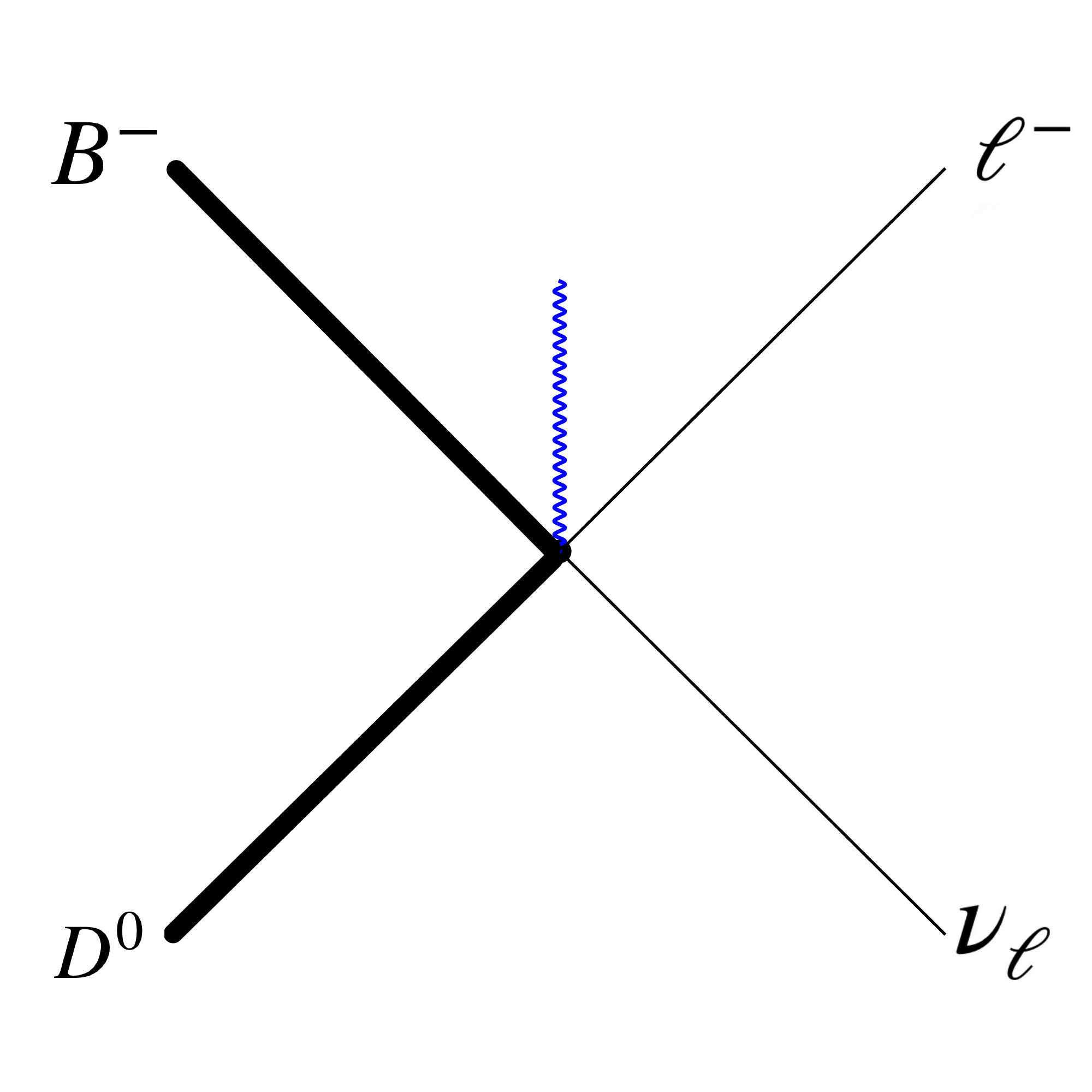}
	 		\caption{}
	 	\end{subfigure}
	 	\caption{Representative diagrams for real photon emission. (a) The photon emission from one of the external charged leg ({\color{red} $\times$} denotes the different possibilities for photon emission). (b) The contact term contribution.}
	 	\label{fig2}
	 \end{figure}

	The Feynman diagrams contributing to the real photon emission are shown in Fig.(\ref{fig2}). Considering the mesons to be point-like and employing scalar QED, the amplitude for $B \rightarrow P \ell \nu_\ell \gamma$, with photon being soft, can be written as a sum of Low's term \cite{PhysRev.110.974} (IR divergent) and 
	an IR safe contribution \cite{Isidori:2007zt}.
	\begin{eqnarray}
	\mathcal{M}_{B\to P \ell \nu_{\ell} \gamma} &=&\mathcal{M}_{\text{IR}}+\mathcal{M}_{\text{NIR}}.\\ 
	\label{eqnm1}
	\text{Here,} \hspace{1.5cm}\mathcal{M}_{\text{IR}}&=&e \epsilon_{\alpha} \mathcal{M}_{0} \left(-\frac{p_{B}^{\alpha}}{p_{B}.k}+\frac{p_{\ell}^{\alpha}}{p_{\ell}.k}\right)
	\end{eqnarray}
	is the Low's soft photon amplitude and the term in paranthesis is called as the universal soft photon function. $\mathcal{M_{\text{NIR}}}$ captures the effect of the internal structure of the mesons in terms of form factors. It consists of the contributions coming from the so called residual term and the contact term (CT).
	The explicit form of $\mathcal{M}_{\text{NIR}}$ is
	\begin{eqnarray}
	\mathcal{M}_{\text{NIR}}&=&\frac{G_{F}}{\sqrt{2}}V_{qb}( \mathcal{M}_{\text{res}} + \mathcal{M}_{\text{CT}}).
	\label{eqnn}
	\end{eqnarray}
	\begin{eqnarray}
	\text{Here,}\nonumber\\
	\mathcal{M}_{\text{res}} &=&e\epsilon_{\alpha}(k)\Big[\Big(\bar{u}(p_{\ell})\gamma^{\alpha}\frac{\slashed{k}}{2p_{\ell}.k}\Gamma^{\mu}v(p_{\nu}) \Big)\otimes H_{\mu}(p_{B},p_{P})+(f_{+}^P+f_{-}^P)\frac{p_{B}^{\alpha}}{p_{B}.k} \bar{u}(p_{\ell})\Gamma^{\mu}v(p_{\nu})\, k_{\mu} \Big],\nonumber\\
	 \mathcal{M}_{\text{CT}}&=& - e\epsilon_{\mu}(k) (f_{+}^P+f_{-}^P) \bar{u}(p_{\ell})\Gamma^{\mu}v(p_{\nu}).
	\end{eqnarray}
	CT is important to ensure gauge invariance of the amplitude and is constructed as in \cite{Mishra:2020orb}. It is important to notice that it is proportional to the charge of the meson and not the lepton which signifies that the leptonic contribution is gauge invariant by itself and CT is necessary to make the hadronic contribution gauge invariant. CT can be introduced via an effective term in the Hamiltonian at the hadronic level, given by	
		\begin{eqnarray}
	\mathcal{H}_{\text{CT}}=-i e(f_+^P-f_-^P) \left[\bar{u}(p_{\ell})\Gamma^{\alpha} v(p_{\nu}) \right]A_{\alpha} \phi^{\dagger}_{P} \phi_{B}.
	\end{eqnarray} 
This term contributes to real as well as virtual corrections. Including the CT contribution, the total gauge invariant amplitude for real soft photon emission reads as
\begin{eqnarray}
	\mathcal{M}_{B\to P \ell \nu_{\ell} \gamma} &=& e \epsilon_{\alpha}(k) \Big[\mathcal{M}_{0} \left(-\frac{p_{B}^{\alpha}}{p_{B}.k}+\frac{p_{\ell}^{\alpha}}{p_{\ell}.k}\right)+\bar{u}(p_{\ell})\frac{\gamma^{\alpha}\slashed{k}}{2p_{\ell}.k}\Gamma_{\mu} v(p_{\nu}) \mathcal{H}^{\mu}\nonumber \\
	&+&  (f_{+}^P+f_{-}^P)\bar{u}(p_{\ell})\left(\frac{p_{B}^{\alpha}}{p_{B}.k}\slashed{k}-\gamma^{\alpha}\right) (1-\gamma^{5})v(p_{\nu}) \Big].
	\end{eqnarray}
	From Eqs. (\ref{eqnm1}) and (\ref{eqnn})
	\begin{eqnarray}
	\left|\mathcal{M}_{B\to P \ell \nu_{\ell} \gamma}\right|^{2} &=&\left|\mathcal{M}_{\text{IR}}\right|^{2}+\left|\mathcal{M}_{\text{res}}\right|^{2} +\left|\mathcal{M}_{\text{CT}}\right|^{2} + 2\mathcal{R}e(\mathcal{M}_{\text{IR}}^{*}\mathcal{M}_{\text{res}})+ 2\mathcal{R}e(\mathcal{M}_{\text{IR}}^{*}\mathcal{M}_{\text{CT}})\nonumber\\ &+& 2\mathcal{R}e(\mathcal{M}_{\text{res}}^{*}\mathcal{M}_{\text{CT}}).\label{2}
	\end{eqnarray}
	Numerically, the contributions from  $\left|\mathcal{M}_{\text{res}}\right|^{2}$, $\left|\mathcal{M}_{\text{CT}}\right|^{2}$, 
	$2\mathcal{R}e(\mathcal{M}_{\text{IR}}^{*}\mathcal{M}_{\text{CT}})$ and 
	$ 2\mathcal{R}e(\mathcal{M}_{\text{res}}^{*}\mathcal{M}_{\text{CT}})$ turn out to be very small 
	(typically contribute at less than 0.1\%). Therefore, we drop these terms and consider only the remaining terms for
	numerical computations which give significant contribution to the decay width. 
	One encounters collinear divergences during these computations. 
	Even though for the case of heavy leptons ($\ell$ being $\mu$ or $\tau$) in the final, the decay rate is less sensitive to 
	collinear divergences, it is important to explicitly check 
	the cancellation of these divergences. For this purpose, it is convenient to consider the photon emission to be inclusive and 
	choose right set of kinematical variables.
	\subsubsection{Photon inclusive}
	 Considering the photon to be inclusive, the total decay width for the process $B\rightarrow P \ell \nu_\ell \gamma $ is
 \begin{eqnarray}
	   \Gamma|_{B\to P \ell \nu_{\ell} \gamma}&=&\frac{1}{2 m_{B}}\int \frac{d^{3}p_{P}}{(2\pi)^{3} 2 E_{P}}\int \frac{d^{3}p_{l}}{(2\pi)^{3} 2 E_{l}}\int \frac{d^{3}p_{\nu}}{(2\pi)^{3} 2E_{\nu}}\int \frac{d^{3}k}{(2\pi)^{3} 2E_{k}}(2\pi)^{4} \delta^{4}\left(Q - p_{\nu}\nonumber\right.\\&-& \left. k \right) \left|\mathcal{M}\right|_{B\to P \ell \nu_{\ell} \gamma}^{2}
	   \label{eqngamma}
	   \end{eqnarray}
	   where $Q=(p_B-p_D-p_\ell)$.\\
	   \begin{figure}[h]
	   	\begin{subfigure}{.5\textwidth}
	   		\centering
	   		\includegraphics[width=0.75\linewidth]{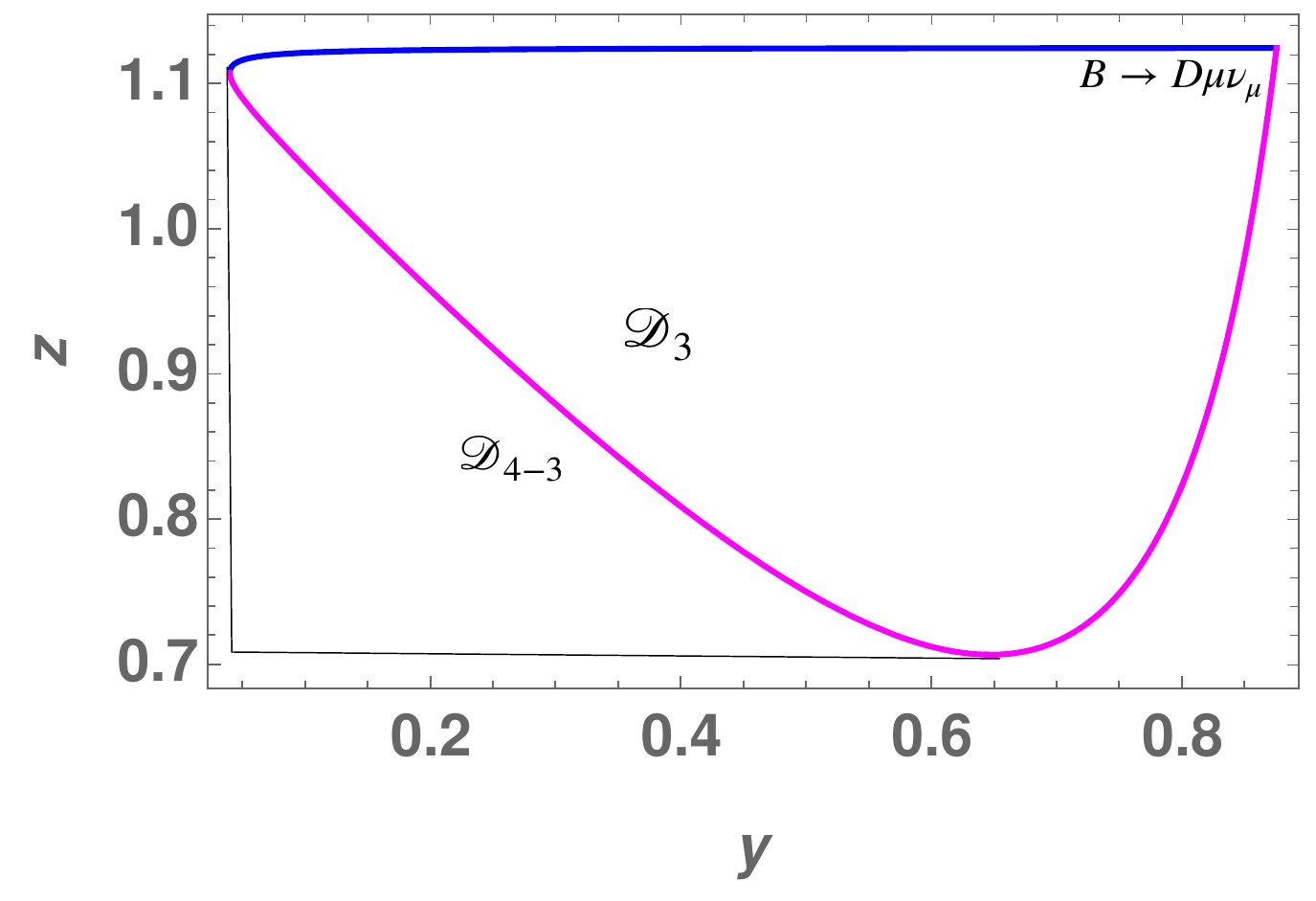}
	   		\caption{}
	   	\end{subfigure}%
	   	\begin{subfigure}{.5\textwidth}
	   		\centering
	   		\includegraphics[width=0.75\linewidth]{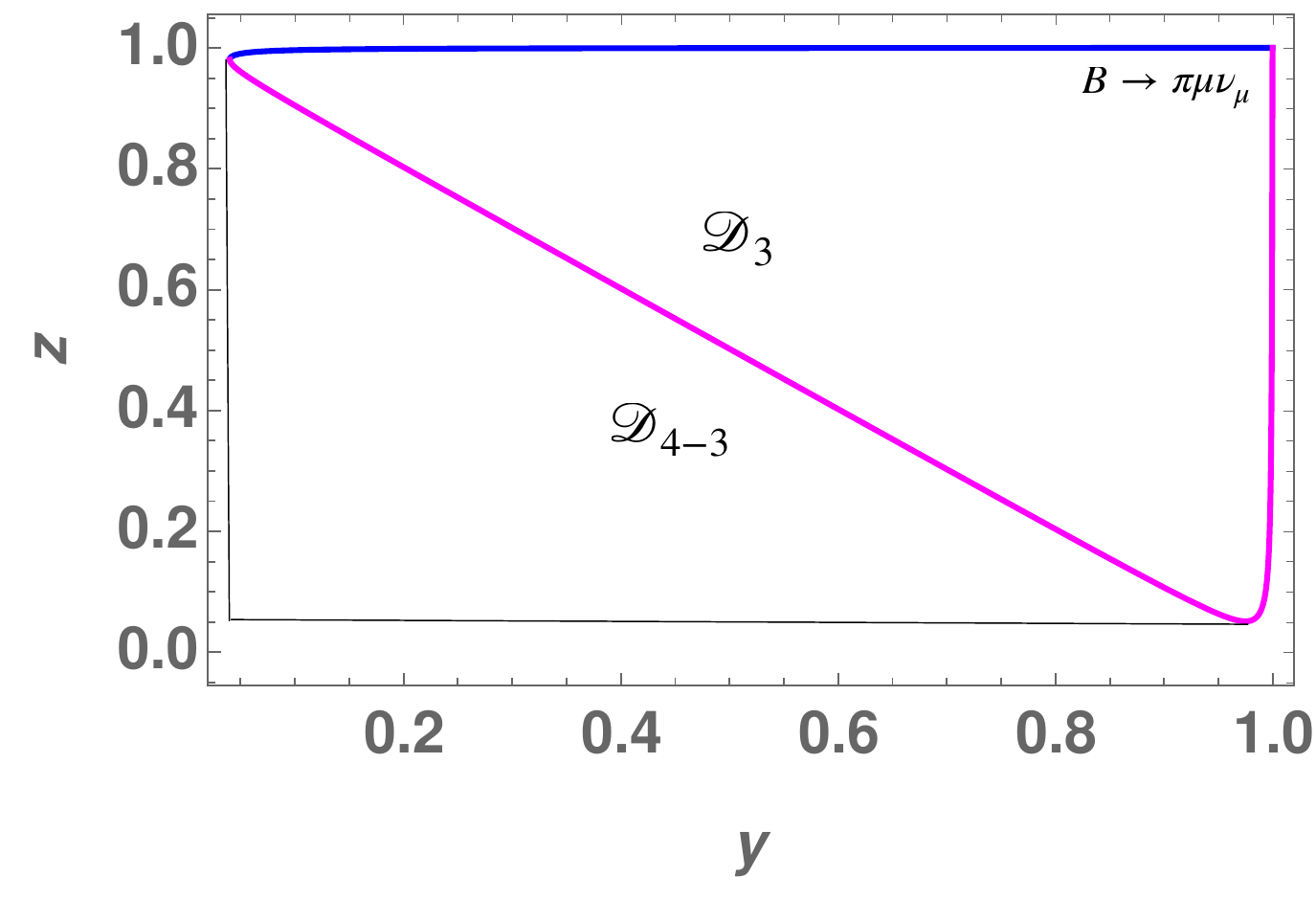}
	   		\caption{}
	   	\end{subfigure}
   	\caption{Dalitz plot showing the phase space boundaries for the lepton (magenta) and meson (blue) energies for the non-radiative processes: (a) $B^- \rightarrow D^0 \mu^- \nu_\mu$ and (b) $B^- \rightarrow \pi^0 \mu^- \nu_\mu$. }
   	\label{fig3}
	   \end{figure} 
	   It is a four body process which contains $B\rightarrow P \ell \nu_\ell$ as a subset. Graphically, it can be seen using the Dalitz plots as shown in Fig. (\ref{fig3}).  The Dalitz plot is found to be linear in the decaying meson's energy and quadratic in the lepton energy.

	It is to be noted here that the delta function present in the Eqn. (\ref{eqngamma}) imposes $x\geq 0$, where $x$ (= $Q^{2}/m_{B}^{2}$) is the normalised total missing mass energy, introduces $\Theta$(x). $\Theta$(x) partitions the full phase space into $\mathcal{D}_3$ (three body region) and $\mathcal{D}_{4-3}$ (the remaining region) which helps us in writing the decay width as
	\begin{eqnarray}
	   \Gamma|_{B\to P \ell \nu_{\ell} \gamma}=&\frac{m_{B}^{3}}{512 \pi^{4}}\Big[\int_{\mathcal{D}_{3}} dy dz \int_{0}^{x_{+}} dx + \int_{\mathcal{D}_{4-3}} dy dz \int_{x_{-}}^{x_{+}} dx \Big] \int \frac{d^{3}k}{(2\pi)^{3} 2E_{k}} \int \frac{d^{3}p_{\nu}}{(2\pi)^{3} 2E_{\nu}}\nonumber \\ &(2\pi)^{4} \delta^{4}(Q - p_{\nu}-k) \left|\mathcal{M}\right|_{B\to P \ell \nu_{\ell} \gamma}^{2}.
	   \label{1}
	   \end{eqnarray}
	   The real photon emission gets contribution from the three body ($\mathcal{D}_3$) as well as the four body ($\mathcal{D}_{4-3}$) phase space regions. Considering the first part of equation ($\ref{1}$) we have  
		{\small \begin{eqnarray}
		\Gamma_{\mathcal{D}_3}|_{B\to P \ell \nu_{\ell} \gamma}&=& \frac{m_{B}^{3}}{512 \pi^{4}}\int_{\mathcal{D}_{3}} dy dz \int_{0}^{x_{+}} dx \int \frac{d^{3}k}{(2\pi)^{3} 2E_{k}}(2\pi)^{4} \delta(x m_{B}^2 - 2 Q.k) \left|\mathcal{M}\right|_{B\to P \ell \nu_{\ell} \gamma}^{2}
		\label{eqnd3}
		\end{eqnarray}}
		\begin{eqnarray}
		\text{with}\hspace{1cm}\left|\mathcal{M}\right|_{B\to P \ell \nu_{\ell} \gamma}^{2}=\left|\mathcal{M}_{\text{IR}}\right|^{2}+ 2\mathcal{R}e(\mathcal{M}_{\text{IR}}^{*}\mathcal{M}_{\text{res}}).
		\end{eqnarray}
		This results in the second order differential decay width (which is free from both IR and collinear divergences once the virtual corrections are also considered) and reads as
		 \begin{eqnarray}
		\frac{d^{2} \Gamma_{\mathcal{D}_3}}{dy dz}=\frac{m_{B}^3}{256\pi^{3}}\frac{\alpha}{\pi}\Big[ \left|\mathcal{M}_{0}\right|^{2} I_{0}(y,z,m_{\gamma}^{2})+\frac{G_{F}^{2}\left|V_{cb}\right|^{2}}{2}\int_{0}^{x_{+}}dx \sum_{m,n} C_{m,n} I_{m,n}(x,y,z) \Big]
		\end{eqnarray}	
		{\small\begin{eqnarray}
		\text{with}\hspace{1cm}I_{m,n} &=& \frac{1}{8\pi}\int \frac{d^{3}p_{\nu}}{E_{\nu}}  \int \frac{d^{3}k}{E_{k}} \delta^{4}(Q- p_{\nu}-k) \frac{1}{(p_{B}.k)^{m} (p_{\ell}.k)^{n}}, \text{  and}\\
         I_{0}&=&\int_{m_{\gamma}^{2}/m_{B}^{2}}^{x_{+}}dx \Big[2 p_{B}.p_{\ell} I_{1,1}(x,y,z) - m_{B}^{2} I_{2,0}(x,y,z) - m_{\ell}^{2} I_{0,2}(x,y,z)\Big].
		\end{eqnarray}}
		The integrals ($I_{0}$, $I_{m,n}$) and the coefficients $C_{m,n}$ are listed in Appendix- \ref{appc}. 
		For practical purposes, it is better to consider the photon exclusive case which is discussed in the next sub-section.
		\subsubsection{Photon exclusive}
    	Now, we consider the exclusive photon case with $k_{max}$ being the maximum energy carried by the soft photon. Following the procedure developed in ref. \cite{Mishra:2020orb}, the second order differential decay width for $B\rightarrow P \ell \nu_\ell \gamma$ (where $\gamma$ is soft) reads as
		\begin{equation}
		\frac{d^{2}\Gamma_{\text{real}}}{dy dz} = \frac{d^{2}\Gamma^{0}}{dy dz}(2 \alpha \tilde{B})  + \frac{d^{2}\Gamma'}{dy dz},
		\label{dr}
		\end{equation}	
		$\frac{d^{2}\Gamma'}{dy dz}$ is IR finite. The IR divergences are contained in $\tilde{B}$ which can be expressed as
		\begin{align}
		\tilde{B}=\frac{-1}{2\pi}\left\lbrace \ln\left( \frac{k_{\text{max}}^{2}m_{B}m_{\ell}}{m_{\gamma}^{2}E_{B}E_{\ell}}\right)-\frac{p_{B}.p_{\ell}}{2}\left[ \int_{-1}^{1}\frac{dt}{p_{t}^{2}}\ln\left(\frac{k_{\text{max}}^{2}}{E_{t}^{2}} \right)+\int_{-1}^{1}\frac{dt}{p_{t}^{2}}\ln\left(\frac{p_{t}^{2}}{m_{\gamma}^{2}} \right) \right]  \right\rbrace.
		\end{align} 
		The overall negative sign in the expression above appears due to charge conservation. $E_t$ and $p_t$ are the combinations of momenta defined as a convenient parametrization to solve the integrals and are given by: $ 
		  2p_{t} = (1+t)p_{B}+(1-t)p_{\ell}$, and 
		   $2E_{t}=(1+t)E_{B}+(1-t)E_{\ell}$, respectively (see Appendix-\ref{appd} for details of the integral). Also, we give photon a small mass ($m_{\gamma}$) which acts as the infrared (IR) regulator.\\
		The $k_{\text{max}}$ dependence of the differential decay width is explicit in this case. As the experiments are unable to report photons of energy smaller than $k_{\text{max}}$, the theoretical rate is expected to depend on $k_{\text{max}}$. Similar to the photon inclusive case, the decay width for the photon exclusive case also contains a non-IR contribution which includes contribution coming from terms beyond Low's term. The terms other than the IR term and its interference with residual term are not significant and hence are not shown explicitly. The interference terms depend on the angle $\theta$ between the lepton and the photon. The angle between the lepton and the neutrino is chosen to be isotropic which leads to $M_{\text{miss}}^2\sim 2E_{\nu}E_\gamma$, where $E_{\nu}=m_B-E_D-E_{\ell}-E_{\gamma}$, in the rest frame of $B$ meson.
	    	\subsection{Virtual Photon Corrections}
	    	\label{subsec2}
\begin{figure}[h]
	    	\begin{subfigure}{.32\textwidth}
	 		\centering
	 		\includegraphics[width=0.45\linewidth]{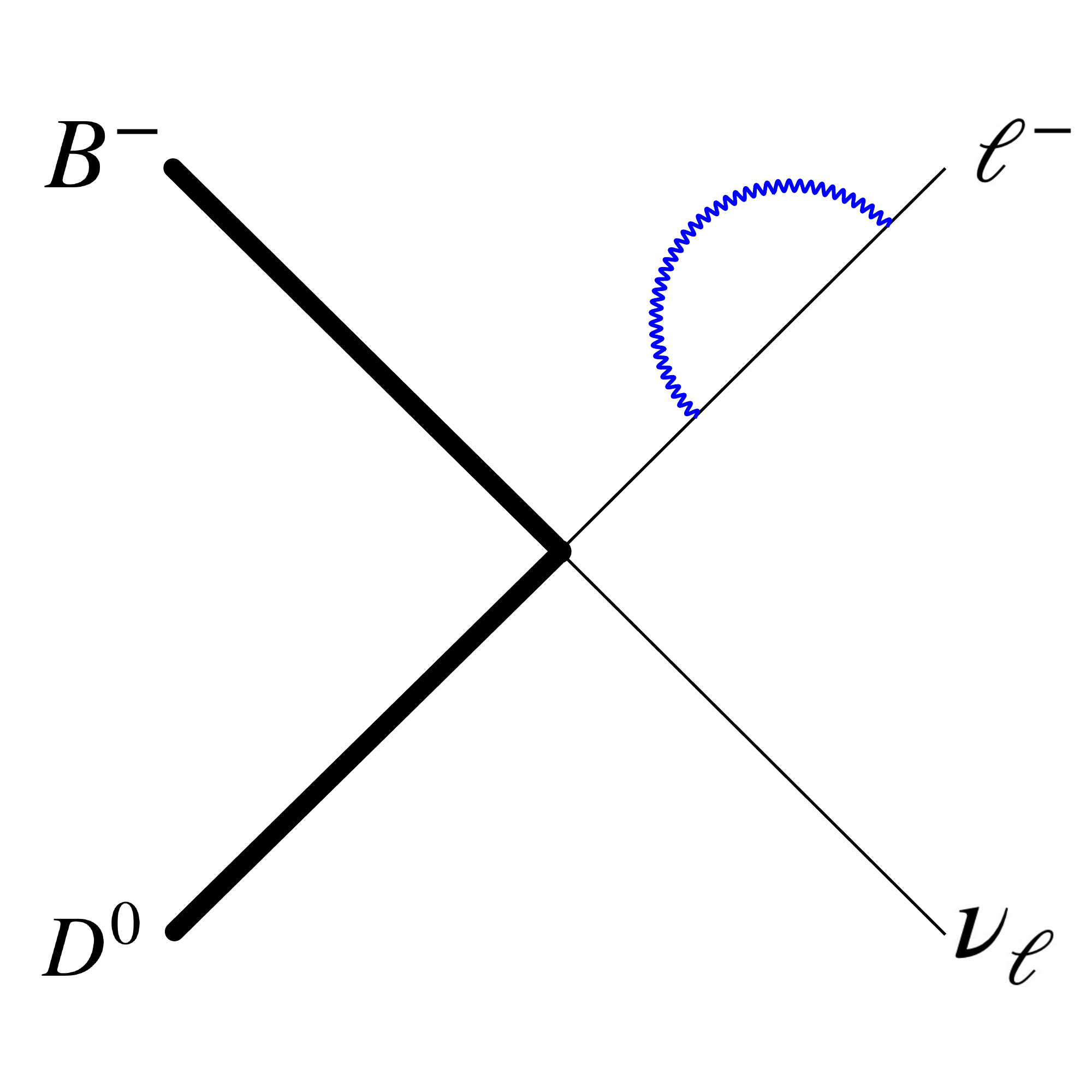}
	 		\caption{}
	 	\end{subfigure}
	 	\begin{subfigure}{.32\textwidth}
	 		\centering
	 		\includegraphics[width=0.45\linewidth]{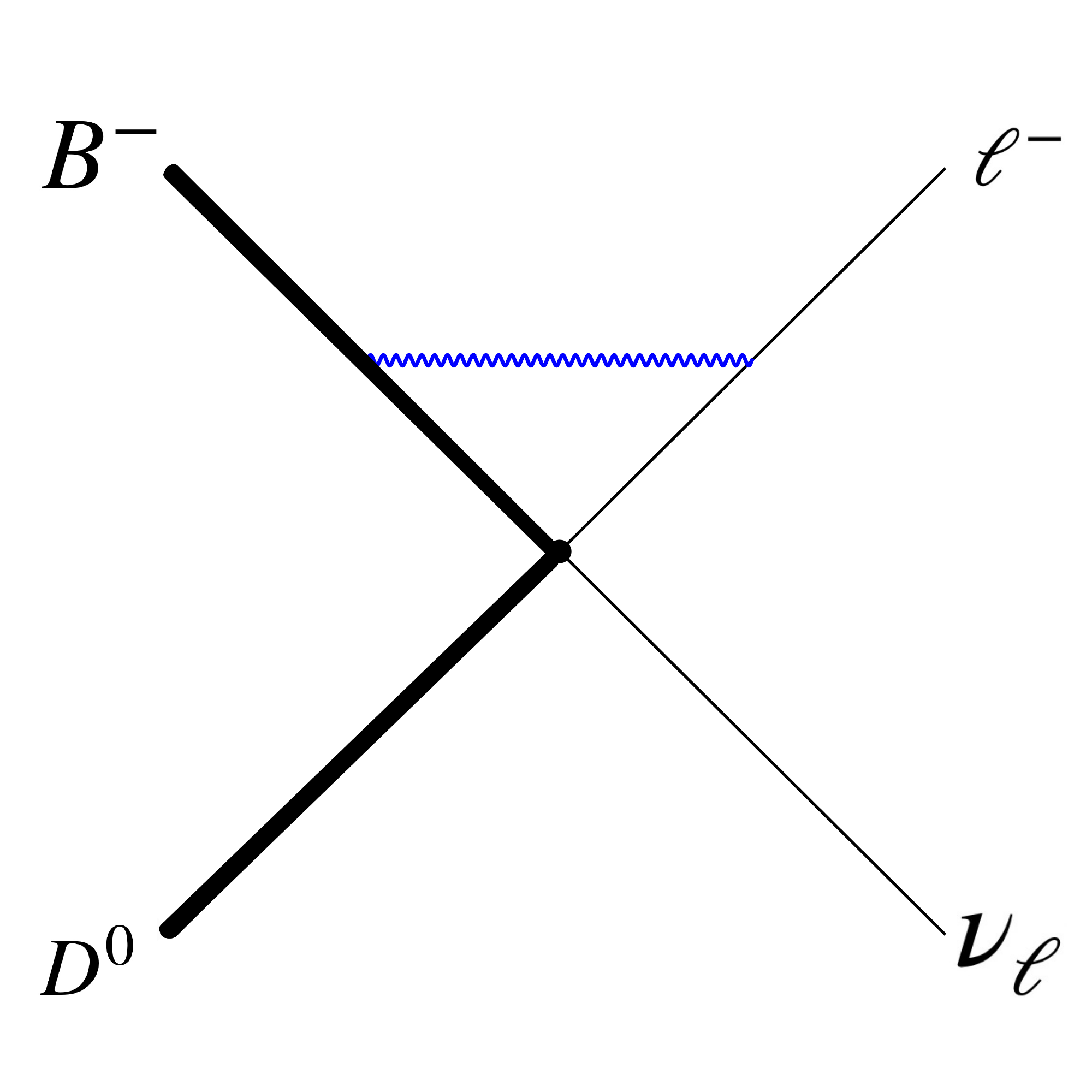}
	 		\caption{}
	 	\end{subfigure}
	    	\begin{subfigure}{.32\textwidth}
	 		\centering
	 		\includegraphics[width=0.45\linewidth]{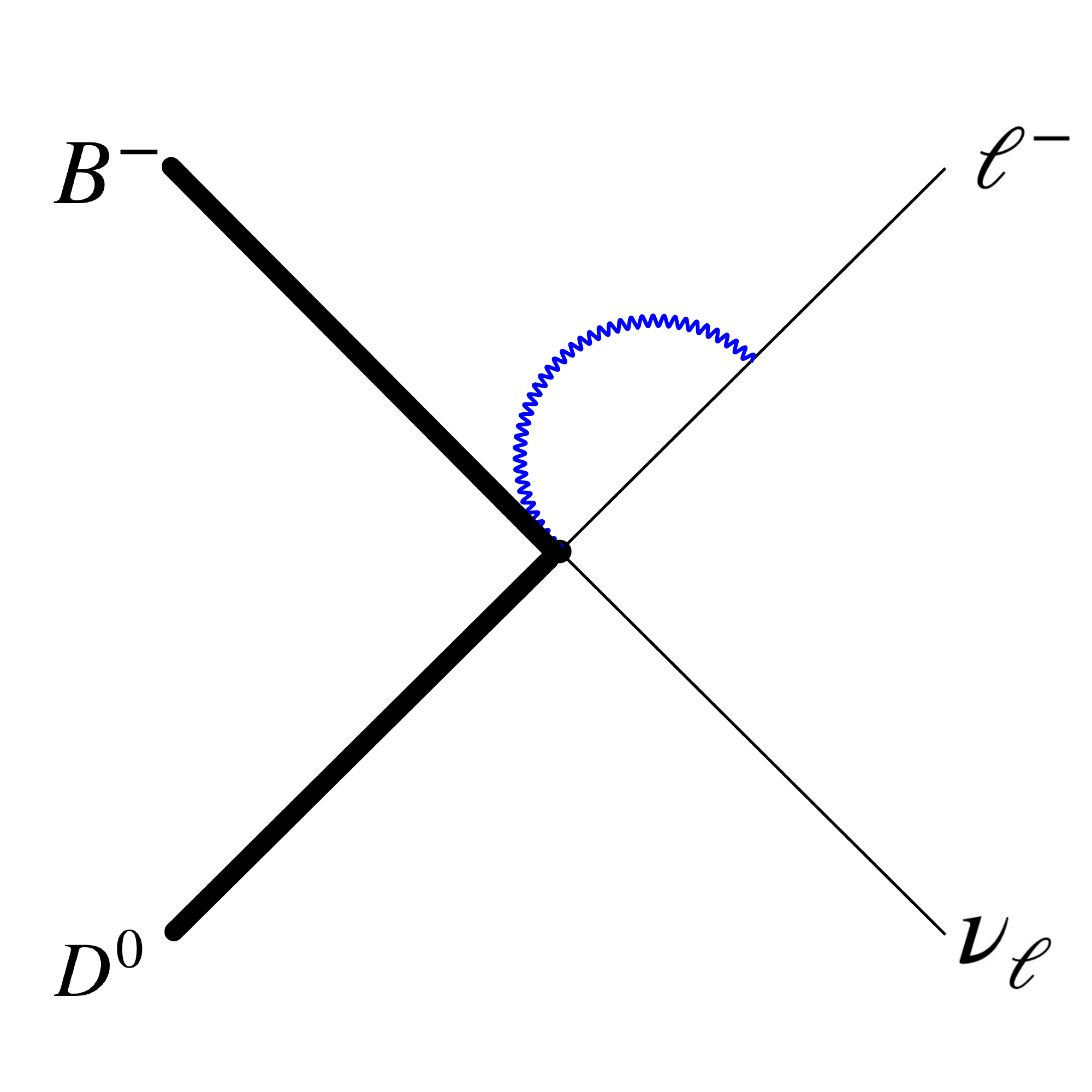}
	 		\caption{}
	 	\end{subfigure}%
	 	\caption{Representative diagrams for virtual photon corrections: (a) the self energy correction to lepton (a similar diagram for B-meson self energy), (b) the vertex correction, and (c) the virtual correction due to contact term (a similar diagram where the photon starts from contact term and ends at B-meson.)}
	 	\label{fig2v}
	 	\end{figure}

	    	There are three types of virtual photon contributions to the process: (1) the self energy correction, where the photon starts and ends at the same charged line (Fig. \ref{fig2v}(a)); (2) the vertex correction, where the photon connects two different charged lines (Fig. \ref{fig2v}(b)); and (3) the contact term contribution, where the photon gets emitted from the effective vertex and ends on a charged particle (Fig. \ref{fig2v}(c)).
	 	These corrections are same for both the photon inclusive as well as the photon exclusive case. Now, we will discuss these three contributions one by one. The contribution due to self energy of the charged lepton and the charged meson (\ref{fig2v}(a)) is given by
	    \begin{eqnarray}
	    	\mathcal{M}_{s}&=& \frac{\mathcal{M}_{0}}{2}(\delta Z_{\ell} + \delta Z_{B})
	    	\end{eqnarray}
	    	where, $\delta Z_{\ell}$ and $\delta Z_{B}$ represents the wave function renormalization of the charged lepton and meson, respectively and are give by
	    	\begin{eqnarray}
	    	\delta Z_{\ell}&=& \frac{\alpha}{4\pi}\Big[2-B_{0}(p_{\ell}^{2},0,m_{\ell}^{2})+4m_{\ell}^{2}B_{0}'(p_{\ell}^{2},m_{\gamma}^{2},m_{\ell}^{2})\Big] , \text{  and} \nonumber\\
	    	\delta Z_{B}&=& \frac{\alpha}{4\pi}\Big[2B_{0}(p_{B}^{2},0,m_{\ell}^{2})+4m_{B}^{2}B_{0}'(p_{B}^{2},m_{\gamma}^{2},m_{B}^{2})\Big].
	    	\end{eqnarray}
	    	Here, $B_{0}(p_{a}^{2},0,m_{a}^{2})$ and $B_{0}'(p_{a}^{2},m_{\gamma}^{2},m_{a}^{2})$ (with $a=\ell (B)$) are the Passarino-Veltman \cite{Passarino:1978jh} functions corresponding to scalar two point integrals and their derivatives (explicit forms are presented in Appendix-\ref{appe}). $B_{0}'(p_{a}^{2},m_{\gamma}^{2},m_{a}^{2})$ contains the IR divergences which are taken care off by providing a fictitious mass, $m_{\gamma}$, to the photon which works as an IR regulator.\\ 
	        The contribution to the matrix element due to vertex correction (as shown in (c) of Fig. \ref{fig2v}) is
	       {\small 
	    \begin{eqnarray}
	    \mathcal{M}_{\text{vert}} &=&  \frac{\alpha}{4\pi} \bar{u}(p_{\ell})\Big[\Big(-2m_{\ell} \slashed{p}_{B}-2\slashed{p}_{B}\slashed{p}_{\ell}\Big)C_{0}(m_{\ell}^{2},m_{B}^{2},q^{2},m_{\ell}^{2},m_{\gamma}^{2},m_{B}^{2}) - \Big(m_{\ell}\slashed{p}_{B} + \slashed{p}_{B}\slashed{p}_{\ell}-2m_{B}^{2}\Big)\nonumber\\ &C_{1}&(m_{B}^{2},q^{2},m_{\ell}^{2},0,m_{B}^{2},m_{\ell}^{2})-\Big(m_{\ell}(\slashed{p}_{\ell}+m_{\ell})+ 2 \slashed{p}_{B}\slashed{p}_{\ell}-4 p_B.p_{\ell}\Big)C_{2}(m_{B}^{2},q^{2},m_{\ell}^{2},0,m_{B}^{2},m_{\ell}^{2})\nonumber\\ &+& B_{0}(q^{2},m_{B}^{2},m_{\ell}^{2})- 2 B_{0}(m_{\ell}^{2},0,m_{\ell}^{2}) \Big]((f_{-}^{p}+f_{+}^{p})\slashed{p}_{\ell} + 2 f_{+}^{p} \slashed{p}_{P})(1-\gamma^{5}) v(p_{\nu}).
	    \end{eqnarray}}
	    Here, $C_{r}(m_{\ell}^{2},m_{B}^{2},q^{2},m_{\ell}^{2},m_{\gamma}^{2},m_{B}^{2})$ ($r=0,1,2$) are the three point Passarino-Veltman functions. $C_{0}$ contains the IR divergences, while the other two functions ($C_1$ and $C_2$) are free from IR divergences, and hence we put $m_{\gamma}^{2}=0$ in these functions. \\
	    The virtual correction due to the CT contributes via the two diagrams: the photon ending on the charged lepton or the charged meson leg. This contribution leads to UV divergences and a finite part. For numerical computations, we discard the UV divergences and incorporate only the finite term. It is found that the finite term contributes very little to the process and hence does not affect the level of precision of the problem. Thus the contact term can be ignored phenomenologically while considering the virtual corrections. \\
	    	To $\mathcal{O}(\alpha)$, the amplitude square for the process $B\rightarrow P \ell \nu_\ell$ including $\mathcal{M}_s$	and $\mathcal{M}_{\text{vert}}$ is 
	    	\begin{eqnarray}
	    	\left|\mathcal{M}\right|^{2}&=& \left|\mathcal{M}_{0}\right|^{2} + 2\mathcal{R}e(\mathcal{M}_{0}^* M_{s}) + 2\mathcal{R}e (\mathcal{M}_{0}^* M_{\text{vert}}) +\mathcal{O}(\alpha^{2})
	    	\end{eqnarray}
	    {\small \begin{eqnarray}
	    	\text{with} \hspace{0.2cm}2\mathcal{R}e(\mathcal{M}_{0}^* M_{s}) &=& \left|\mathcal{M}_{0}\right|^{2}(\delta Z_{\ell} + \delta Z_{B}), \hspace{0.5cm}\text{and} \nonumber\\
	    		2\mathcal{R}e (\mathcal{M}_{0}^* M_{\text{vert}})&=&\frac{\alpha}{4\pi}\Big[\left|\mathcal{M}_{0}\right|^{2}\Big(2  B_{0}(q^{2},m_{B}^{2},m_{\ell}^{2}) - 4  B_{0}(m_{\ell}^{2},0,m_{\ell}^{2})-4\left((p_{B}.p_{\ell})+ m_{B}^{2}\right)\nonumber\\&&C_{1}(m_{B}^{2},q^{2},m_{\ell}^{2},0,m_{B}^{2},m_{\ell}^{2})- 8(p_{B}.p_{\ell})  C_{0}(m_{\ell}^{2},m_{B}^{2},q^{2},m_{\ell}^{2},m_{\gamma}^{2},m_{B}^{2})- 4 m_{\ell}^{2}\nonumber\\&& C_{2}(m_{B}^{2},q^{2},m_{\ell}^{2},0,m_{B}^{2},m_{\ell}^{2})\Big)+\Big((-4 f_{+}^{p}(f_{-}^{p}+f_{+}^{p})-2(f_{-}^{p}+f_{+}^{p})^{2})(p_{B}.p_{\ell})(p_{\ell}.p_{\nu})\nonumber\\&+&4 (f_{+}^{p})^{2}m_{P}^{2}(p_{B}.p_{\nu})+(p_{P}.p_{\nu})(-4(f_{-}^{p}+f_{+}^{p})f_{+}^{p}-4f_{+}^{p} (p_{B}.p_{P}) ) + 4f_{-}^{p}f_{+}^{p} (p_{B}.p_{\nu}) (p_{\ell}.p_{P}) \nonumber\\&+& (f_{+}^{p}+f_{-}^{p})^{2}m_{\ell}^{2} (p_{B}.p_{\nu}) + 4(f_{+}^{p})^{2}(p_{B}.p_{\nu}) (p_{\ell}.p_{P}) \Big)C_{2}(m_{B}^{2},q^{2},m_{\ell}^{2},0,m_{B}^{2},m_{\ell}^{2})\Big],
	    	\label{eqnmat}
	    	\end{eqnarray}}
respectively. Therefore, the differential non-radiative decay width including the virtual QED corrections reads as
	    	\begin{eqnarray}
	    	\frac{d^{2} \Gamma_{\text{vir}}}{dy dz}= \frac{d^{2} \Gamma^{0}}{dy dz}(1+2 \alpha B)+ \frac{d^{2} \Gamma'_{\text{vir}}}{dy dz}. 
	    	\label{dv}
	    	\end{eqnarray}
Here, $ \frac{d^{2} \Gamma'_{\text{vir}}}{dy dz}$ is IR-finite and contains the corrections due to non-factorizable terms (combination of form factors 
and momenta) present in Eq. (\ref{eqnmat}). The factorizable correction factor, $B$, is IR divergent and reads as
{\small
	    	\begin{eqnarray}
	    	B&=&\frac{1}{8\pi}\Big[2 B_{0}(q^{2},m_{B}^{2},m_{\ell}^{2})- 4  B_{0}(m_{\ell}^{2},0,m_{\ell}^{2}) -4\left((p_{B}.p_{\ell})+ m_{B}^{2}\right)C_{1}(m_{B}^{2},q^{2},m_{\ell}^{2},0,m_{B}^{2},m_{\ell}^{2})\nonumber\\&-&8(p_{B}.p_{\ell})  C_{0}(m_{\ell}^{2},m_{B}^{2},q^{2},m_{\ell}^{2},m_{\gamma}^{2},m_{B}^{2})-4 m_{\ell}^{2}C_{2}(m_{B}^{2},q^{2},m_{\ell}^{2},0,m_{B}^{2},m_{\ell}^{2})+2-B_{0}(p_{\ell}^{2},0,m_{\ell}^{2})\nonumber\\&+&4m_{\ell}^{2}B_{0}'(p_{\ell}^{2},m_{\gamma}^{2},m_{\ell}^{2})+2B_{0}(p_{B}^{2},0,m_{\ell}^{2})+4m_{B}^{2}B_{0}'(p_{B}^{2},m_{\gamma}^{2},m_{B}^{2})\Big].
	    	\end{eqnarray}}
	  	\subsection{Total $\mathcal{O}(\alpha)$ QED corrections}
	  	\label{subsec3}
	    	After summing $\frac{d^{2} \Gamma_{\text{real}}}{dy dz}$ and $\frac{d^{2} \Gamma_{\text{vir}}}{dy dz}$, the double differential decay width for the process $B \rightarrow P \ell \nu_\ell$, including real and virtual soft photon corrections can be written at $\mathcal{O}(\alpha)$ as,
	    		\begin{eqnarray}
		 	\frac{d^{2}\Gamma_{\ell}^{\text{QED}}}{dy dz}=\frac{d^{2}\Gamma^{0}}{dy dz}\left(1 + 2\alpha \mathcal{H}\right)+ \frac{d^{2}\Gamma'}{dy dz}+\frac{d^{2} \Gamma'_{\text{vir}}}{dy dz},
		 	\label{totalqed}
		 	\end{eqnarray}
		 	where $\mathcal{H}=\tilde{B}+B$. Though, $\tilde{B}$ and $B$ depend on IR regulator $m_{\gamma}$, there sum, $\mathcal{H}$ is independent of $m_{\gamma}$. Hence, the IR divergences cancel in the sum. The explicit of form of $\mathcal{H}$ is 
		 	\begin{eqnarray}
		 	\mathcal{H}&=&\frac{1}{2\pi}\Bigg[ -\ln\left(\frac{k_{max}^2}{E_B E_{\ell}}\right)+\frac{p_B.p_{\ell}}{2}\int_{-1}^{1} \frac{dt}{p_t^2}\frac{k_{max}^2}{E_t^2}+ B_{0}(q^{2},m_{B}^{2},m_{\ell}^{2})- 2  B_{0}(m_{\ell}^{2},0,m_{\ell}^{2}) \nonumber\\&-&2\left((p_{B}.p_{\ell})+ m_{B}^{2}\right)C_{1}(m_{B}^{2},q^{2},m_{\ell}^{2},0,m_{B}^{2},m_{\ell}^{2}) -2 m_{\ell}^{2}C_{2}(m_{B}^{2},q^{2},m_{\ell}^{2},0,m_{B}^{2},m_{\ell}^{2})\nonumber\\&-&3-\frac{1}{2}B_{0}(p_{\ell}^{2},0,m_{\ell}^{2})+B_{0}(p_{B}^{2},0,m_{B}^{2})\Big].
		 	\label{H}
		 	\end{eqnarray}
		 	We like to recall here that the terms $\frac{d^{2}\Gamma'}{dy dz}$ and $\frac{d^{2} \Gamma'_{\text{vir}}}{dy dz}$ in Eqn. (\ref{totalqed}) are free from IR divergences. Hence, $\frac{d^{2}\Gamma_{\ell}^{\text{QED}}}{dy dz}$ is an IR safe quantity and can be written in the compact as
	        \begin{equation}
	    	\frac{d^{2}\Gamma_{\ell}^{\text{QED}}}{dy dz}=\frac{d^{2}\Gamma^{0}}{dy dz}\left(1 + \Delta_{\ell}^{\text{QED}} \right) .
	    	\label{delta}
	    	\end{equation}
	    Here, $\ell=\mu$, $\tau$ and $\Delta_{\ell}^{\text{QED}}$ is the correction factor to the decay width. It contains corrections due to infrared and non-infrared factors up to $\mathcal{O}(k)$. The $\mathcal{O}(k^{2})$ term was explicitly checked to be small and hence has been ignored in the numerical analysis.\\
	    		    	Following Eq.(\ref{nondefde}), the CKM element $|V_{qb}|$ without inclusion of QED corrections can be written as
	    	\begin{eqnarray}
	   	|V_{qb}^{0}|=\sqrt{\frac{\Gamma_{qb}^{\text{exp}}}{\mathcal{G}_{qb}^{0}}}.
	   	\label{eqnvqb}
	   	\end{eqnarray}
	   	Here, $\Gamma^{\text{exp}}_{qb}$ is experimental decay width, and $\mathcal{G}_{qb}^0$ is defined as
	   	\begin{eqnarray}
\mathcal{G}_{qb}^0 	=\frac{m_{B}}{256\pi^{3}}\frac{G_{F}^{2}}{2}\int dy \int dz |\mathcal{M}_{0}|^{2} \hspace{0.5cm} (q=u/c).
	\end{eqnarray}
	Therefore the ratio of the CKM elements without QED corrections, defined as $R_V^0$, will be
	\begin{eqnarray}
	   	R_{V}^0=\frac{|V_{ub}^0|}{|V_{cb}^0|}=\sqrt{\frac{\Gamma^{0}_{ub}\mathcal{G}_{cb}^0}{\Gamma^{0}_{cb}\mathcal{G}_{ub}^0}}.
	   	\label{vuc}
	   	\end{eqnarray}
	   	Taus are harder to recontruct while electrons are far more sensitive to soft photon corrections.
	   	Therefore, for extracting the CKM elements and their ratio, it is advisable to choose final states with muons. 
	   	As only muons are considered in the final states, the collinear logs $\sim \ln(m_{\mu})$
are the same for both $B\to \pi$ and $B\to D$ transitions. The QED correction factors for $V_{qb}$ and $R_V$ are defined as,
	\begin{eqnarray}
	\delta_{V_{qb}}^{\text{QED}}=\frac{|V_{qb}|}{|V_{qb}^0|}-1,\hspace{0.5 cm}\text{and}\\
	   	\Delta_{R_{V}}= \delta_{V_{ub}}^{\text{QED}}-\delta_{V_{cb}}^{\text{QED}}
	   	\label{deltaRV},
	   	\end{eqnarray}
	   	respectively. Here $|V_{qb}|$ is the CKM element including the QED corrections to $\mathcal{O}(\alpha)$.\\
    For completeness, We also consider the soft photon corrections to the ratio $R_P$ (i.e. the ratio of branching fraction of $\tau$ mode to $\mu$ mode), and is given by 
	\begin{eqnarray}
	\delta_{R_{P}}=R_P^0\Big(\frac{\Delta_\tau^{\text{QED}}}{\Gamma_{\tau}^0}-\frac{\Delta_\mu^{\text{QED}}}{\Gamma_{\mu}^0}\Big)
	\end{eqnarray}
	 where, $\Delta_\tau^{\text{QED}}$ and $\Delta_\mu^{\text{QED}}$ are the correction factors in the $\tau$ and $\mu$ mode, respectively.

    \section{Results}
	  \label{sec4}
	The soft photon correction to the process $B\rightarrow P \ell\nu_\ell$ is studied. The experimental analyses follow two 
	approaches to study $B\rightarrow P \ell\nu_\ell(\gamma)$: ($1$) the photon inclusive approach, where only the charged mesons and 
	leptons are detected while the neutrino and the photon are left undetected;
	($2$) The photon exclusive approach, where the experiment is sensitive to the final state radiated photons. For the inclusive case,
	the observed momenta of the charged mesons and leptons are fitted to the three body kinematics with zero or non-zero missing mass.
	The full decay width turns out to be a function of the maximum and minimum of the missing mass. In this work, we focused on 
	the exclusive case as we are interested in the study of the explicit effect of soft photons on the process. 
	As the photon will now be detected, a fraction of the four body phase space (upto $k_{max}$) will also contribute.
	Hence, the total decay width gets contribution from the three-body phase space region as well as beyond it. \\
	 The real emission of the soft photon contributes to inside as well as outside of the Dalitz region
	 (shown in \ref{fig3}(a) and \ref{fig3}(b)). While the virtual correction contributes only to inside of the Dalitz region.
The correction factor turns out to be less sensitive to phase space points outside the Dalitz region as the leptons at hand are heavy.
But this region is important to see the long distance effects ($k\to 0$) and to enhance the precision. It also helps to see the dependence of
the decay width on the angle between photon and $\ell$, $\theta$ which dictates the collinear divergences. We found that the correction factor due to soft
photon denoted by $\Delta_{\ell}^{QED}$ ($\ell=\mu$ and $\tau$) is not much sensitive to the cut on $\theta$
(denoted by $\theta_{cut}$) for both muons and taus because of their large mass. 
Had we been considering electrons in the final state, this dependence would have been significant. 
 Broadly, the correction factor $\Delta_{\ell}^{\text{QED}}$ is found to be more for the muon channel (roughly 3-5 \%) as compared to the tau channel (almost negligible) for both $B\to D$ and $B\to \pi$ decay modes. To be explicit, (say, at $k_{max}= 100$ MeV), $B^-\to D^0$ ($B^0\to D^+$) mode receives QED shift of $\sim 0.1\%$($\sim -1\%$) for $\tau$ mode and $\sim -1.6\%$ ($\sim -3.4\%$) for $\mu$ mode.\\ 
	  \begin{figure}[h]
	 	\begin{subfigure}{.5\textwidth}
	 		\centering
	 		\includegraphics[width=0.9\linewidth]{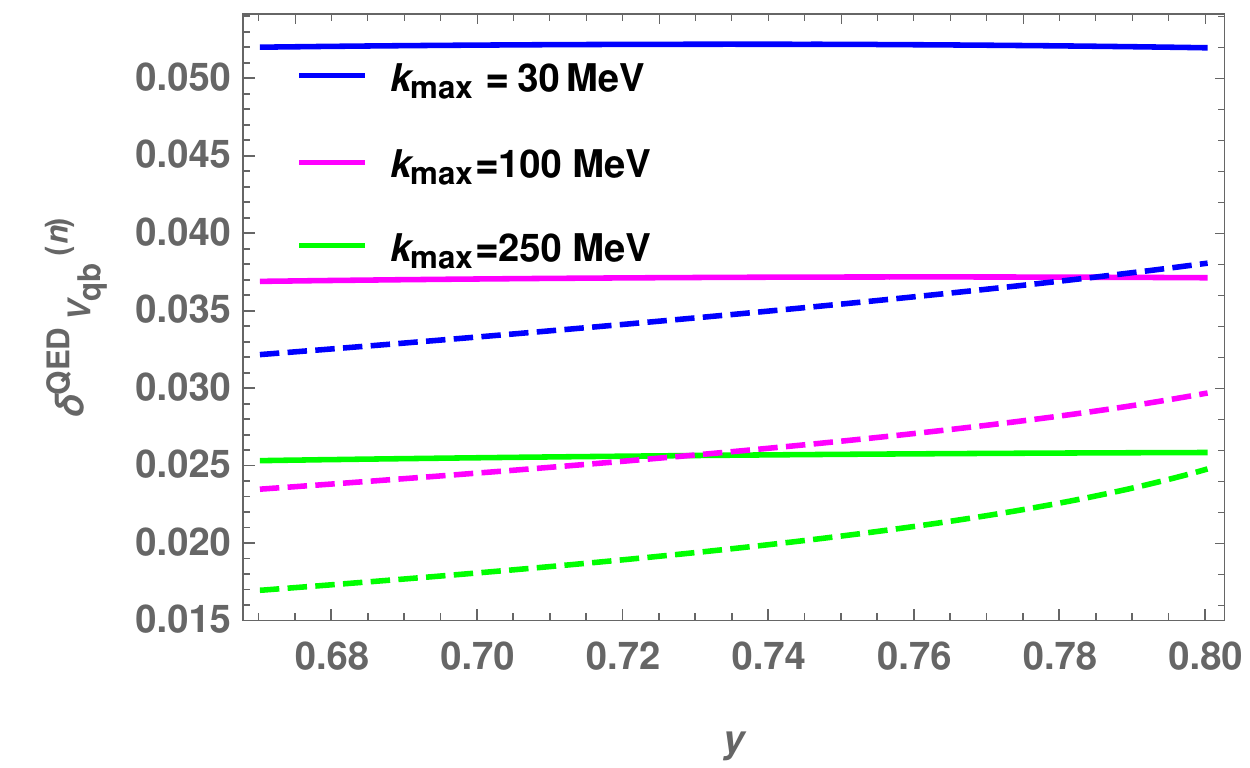}
	 		\caption{}
	 	\end{subfigure}%
	 	\begin{subfigure}{.5\textwidth}
	 		\centering
	 		\includegraphics[width=0.9\linewidth]{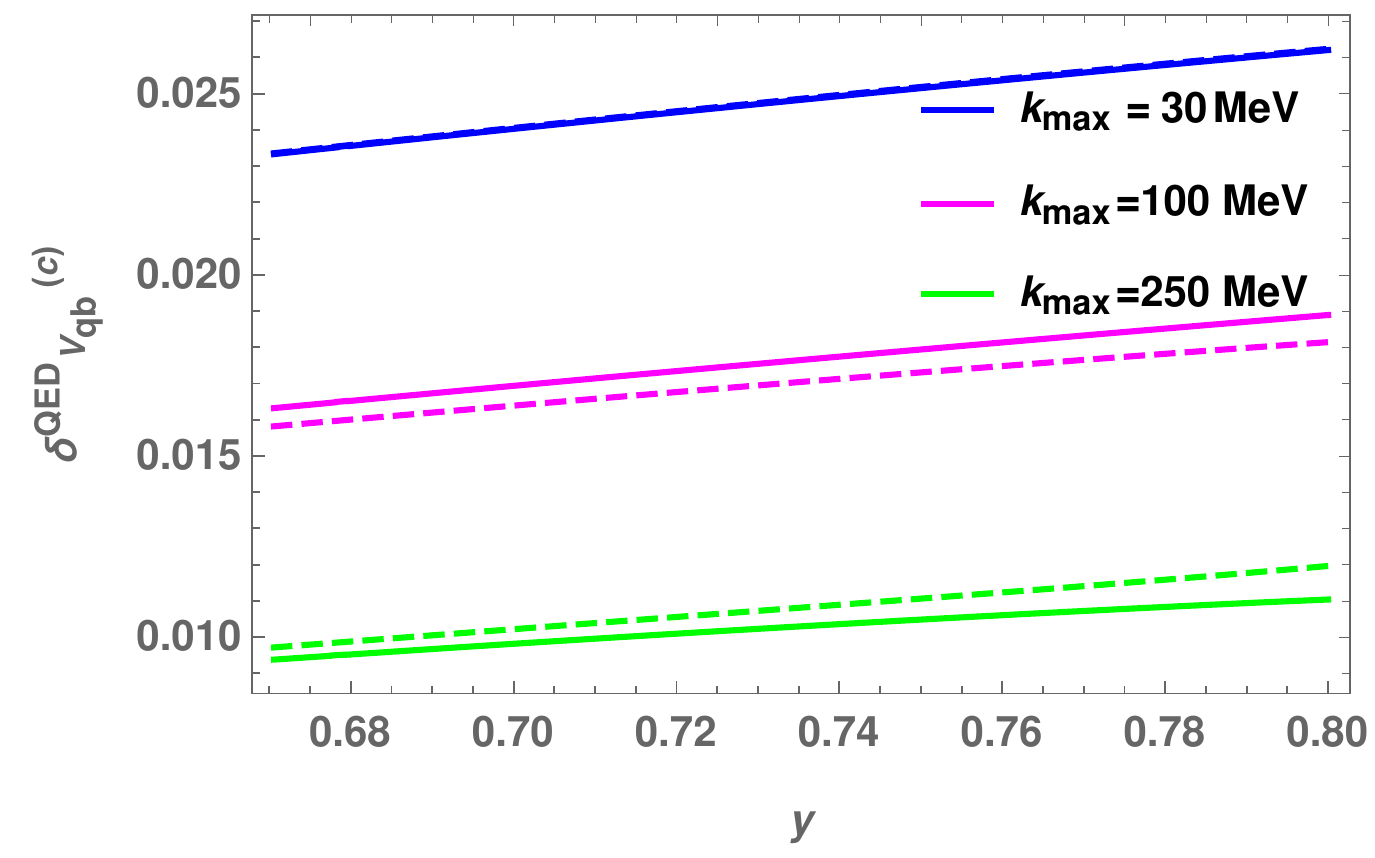}
	 		\caption{}
	 	\end{subfigure}
	 	\caption{Radiative corrections to the CKM elements $|V_{cb}|$ and $|V_{ub}|$ (i.e. $\delta^{\text{QED}}_{V_{cb}}$ (dashed) and $\delta^{\text{QED}}_{V_{ub}}$ (solid) for different  thresholds on photon energy, $k_{max}$ for (a) $B^0 \rightarrow P^+(=D^+,\pi^+) \mu^- \nu_\mu$ and (b) $B^- \rightarrow P^0(=D^0,\pi^0) \mu^- \nu_\mu$.}
	 	\label{fig5}
	 \end{figure} 
	 Figures \ref{fig5}(a) and \ref{fig5}(b) illustrate the soft photon corrections to the CKM elements $|V_{cb}|$ and $|V_{ub}|$ for neutral and charged modes, respectively. The corrections to both the CKM elements when considering the charged mode are found to be almost same, as the photon is getting emitted from the $B$-meson and the lepton in both $B\rightarrow \pi \ell \nu_\ell$ and $B \rightarrow D \ell \nu_\ell$. However, for the neutral mode, we observe some difference between the two curves as now the photon is getting emitted from $\pi$ and $D$ instead of B, hence their mass difference plays a crucial role.\\
	 Next, we study the effect of soft photons on the ratio of these CKM elements defined by $R_{V} = \frac{V_{ub}}{V_{cb}}$. 
	 The correction factor to this ratio, $\Delta_{R_{V}}$ as a function of lepton energy is shown in Fig. (\ref{fig6}) for both the 
	 neutral as well as charged modes. It is found that the charged mode gets almost zero correction while the neutral mode gets	very minute ($\sim\mathcal{O}(10^{-3})$) correction due to soft photons. This difference emerges as a consequence of photon emission from $\pi$ vs $D$ in the neutral case as discussed above. We have also studied the dependence of the correction factor on the choice of maximum energy of photon, $k_{\text{max}}$. It is found that the correction factor decreases with an increase in $k_{max}$ as the collinear and IR effects are more and more suppressed (can be seen from Fig. (\ref{fig6})), similar trends can be seen in Ref. \cite{deBoer:2018ipi}.
	 	\begin{table}[h]
	   \centering
    \begin{tabular}{ |c| c| c|}
    \hline
    {  }  & 
    \begin{tabular}{@{}c@{}}$\Delta_{\tau}^{(n)}$ \\  {(with (w/o) Coulomb)}\end{tabular}
    & 
    \begin{tabular}{@{}c@{}}$\Delta_{\mu}^{(n)}$\\  {(with (w/o) Coulomb)}\end{tabular}
    \\
    \hline
    {Ref. \cite{deBoer:2018ipi}}& {$1.7(-1.2)$} & {$-1.2(-3.5)$}   \\
    \hline
     {Our results}& {$1.7(-1.0)$} & {$-1.1(-3.4)$}    \\
    \hline
    \end{tabular}\\
	    \caption{ Comparison of numerical values of QED shifts (\%) in the decay width (for both tau and $\mu$ modes) at $k_{max}=100$ MeV with Ref. \cite{deBoer:2018ipi} considering (not-considering) the extra Coulomb factor from the final state charged particles (as considered in Ref. \cite{deBoer:2018ipi}).}
	    \label{tablecomp}
	\end{table}
	For the neutral $B$ mode Ref. \cite{deBoer:2018ipi} has included an additional coulomb factor coming from the final state charged particles. We have found our results to be in full agreement with the findings of Ref. \cite{deBoer:2018ipi} (see Table ($\ref{tablecomp}$) for a comparative study). 
	 \begin{figure}[h]
	 		\centering
	 		\includegraphics[width=0.5\linewidth]{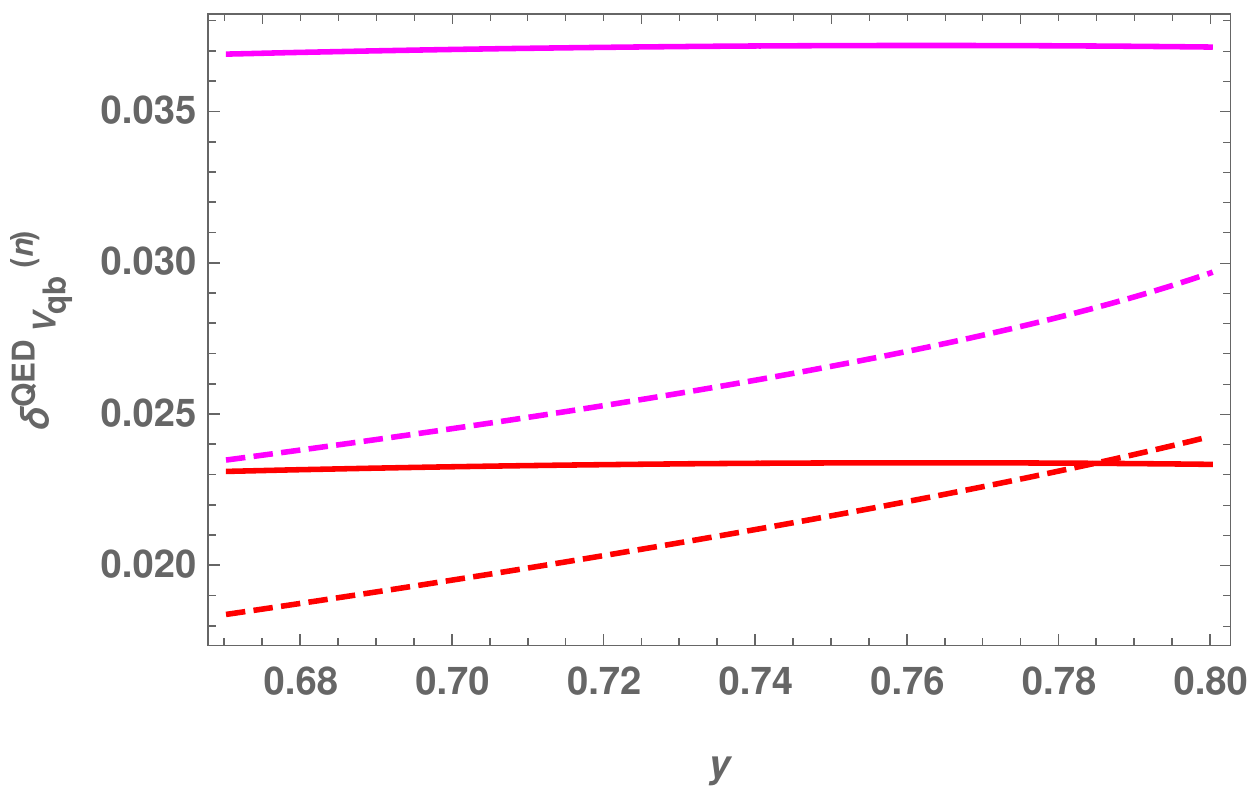}
     	\caption{Radiative corrections to the CKM elements $|V_{cb}|$ and $|V_{ub}|$ (i.e. $\delta^{\text{QED}}_{V_{cb}}$ (dashed) and $\delta^{\text{QED}}_{V_{ub}}$ (solid)) for $k_{max}=100$ MeV for $B^0 \rightarrow P^+(=D^+,\pi^+) \mu^- \nu_\mu$ considering (Red) and not-considering (Magenta) the extra Coulomb factor from the final state charged particles.}
	 	\label{figcomp}
	 \end{figure}
	The QED corrected CKM elements $|V_{cb}|$ and $|V_{ub}|$ for  neutral $B$ mode with and without this extra Coulomb factor are shown in Fig. (\ref{figcomp}). This factor reduces the QED effects from $\sim 3\%$ to $\sim 2\%$. We would like to emphasize here that, though this factor has an impact on QED corrections to the individual CKM elements, it has negligible effect on the proposed observable $R_V$, demonstrating its robustness against all variety of QED corrections.
	  \begin{figure}[h]
     \begin{subfigure}{.5\textwidth}
	 		\centering
	 		\includegraphics[width=0.9\linewidth]{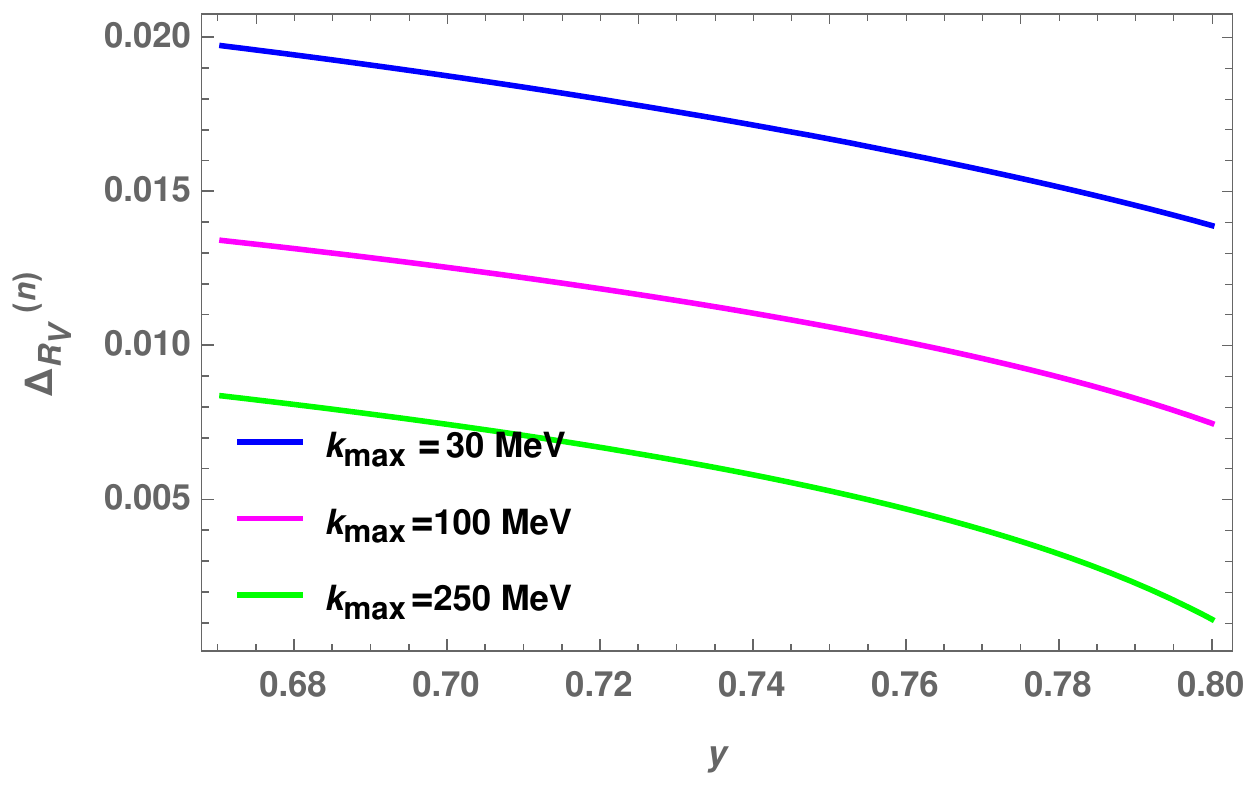}
	 		\caption{}
	 	\end{subfigure}
     \begin{subfigure}{.5\textwidth}
	 		\centering
	 		\includegraphics[width=0.9\linewidth]{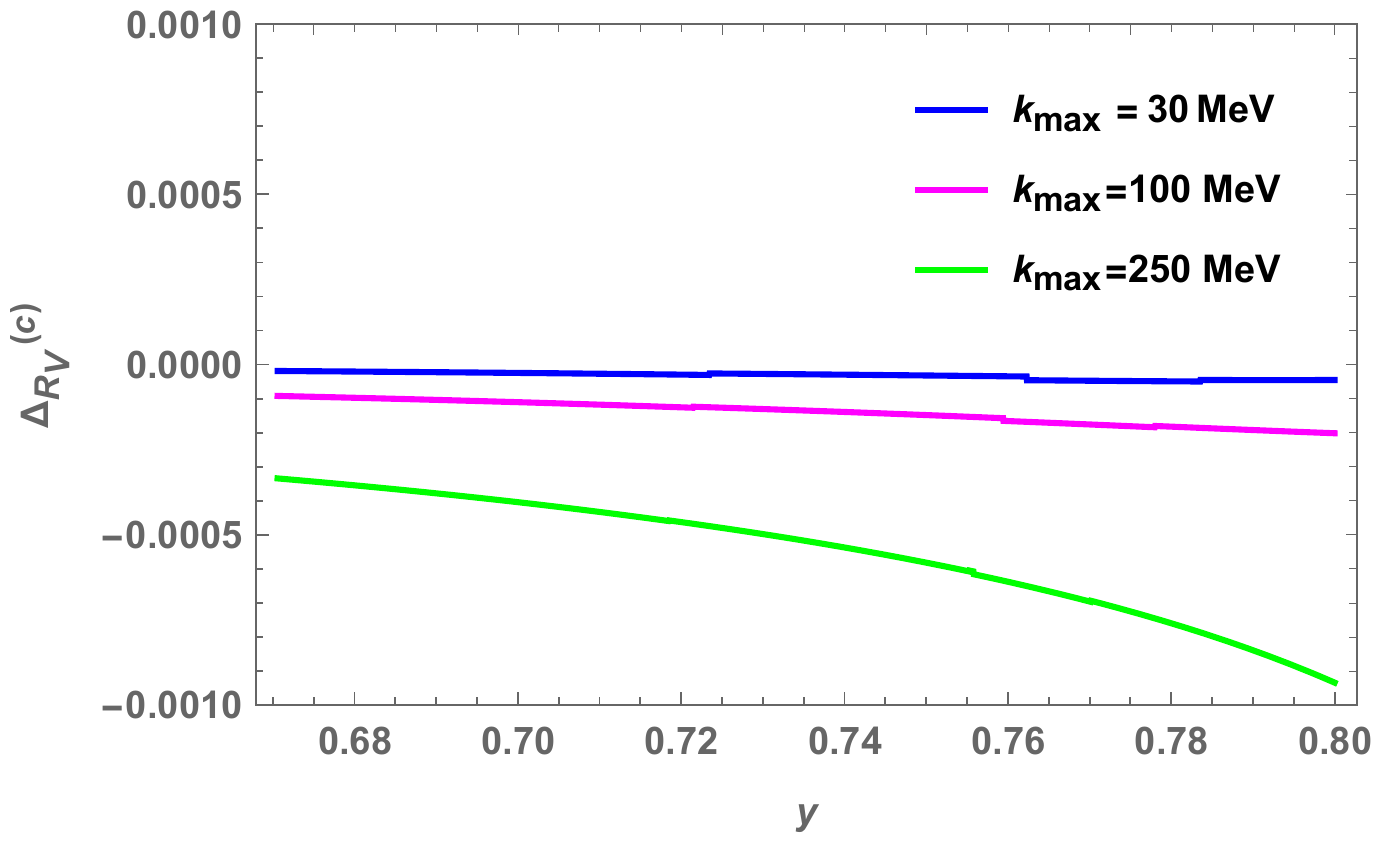}
	 		\caption{}
	 	\end{subfigure}%
     	\caption{Radiative corrections to $V_{ub}/V_{cb}$ (i.e. $\Delta{R_V}$) for different thresholds on photon energy, $k_{max}$ for (a) $B^0 \rightarrow P^+(=D^+,\pi^+) \mu^- \nu_\mu$ and (b) $B^- \rightarrow P^0(=D^0,\pi^0) \mu^- \nu_\mu$.}
	 	\label{fig6}
	 \end{figure} 
	\begin{table}[h]
	   \centering
    \begin{tabular}{ |c| c| c|c|c|}
    \hline
    {  }  & {$(f_{B\to\pi}^{(I)};f_{B\to D}^{(I)})$} & {$(f_{B\to \pi}^{(II)};f_{B\to D}^{(I)})$} & {$(f_{B\to\pi}^{(I)};f_{B\to D}^{(II)})$} &
    {$(f_{B\to \pi}^{(II)};f_{B\to D}^{(II)})$}\\
    \hline
    {$R_V$}& {$0.091$} & {$0.093$} & {$0.091$} & {$0.093$}    \\
    \hline
    \end{tabular}\\
    \label{table2}
	    \caption{ The ratio of $R_V$ determined with the choice $f_{B\to \pi}^{(A)}$ and $f_{B\to D}^{(A)}$ for the corresponding form factors.}
	    \label{tabel3}
	\end{table}

\begin{figure}[h]
	 	\begin{subfigure}{.5\textwidth}
	 		\centering
	 		\includegraphics[width=0.9\linewidth]{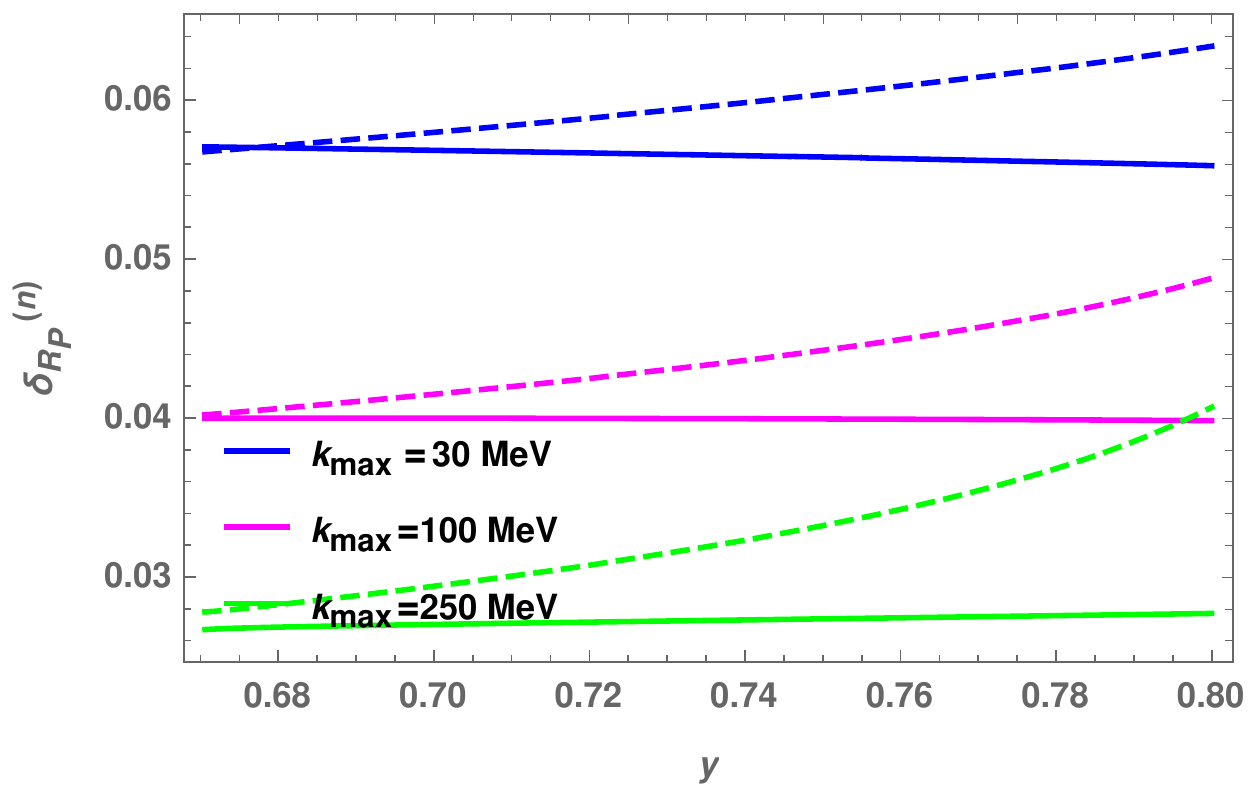}
	 		\caption{}
	 	\end{subfigure}%
	 	\begin{subfigure}{.5\textwidth}
	 		\centering
	 		\includegraphics[width=0.9\linewidth]{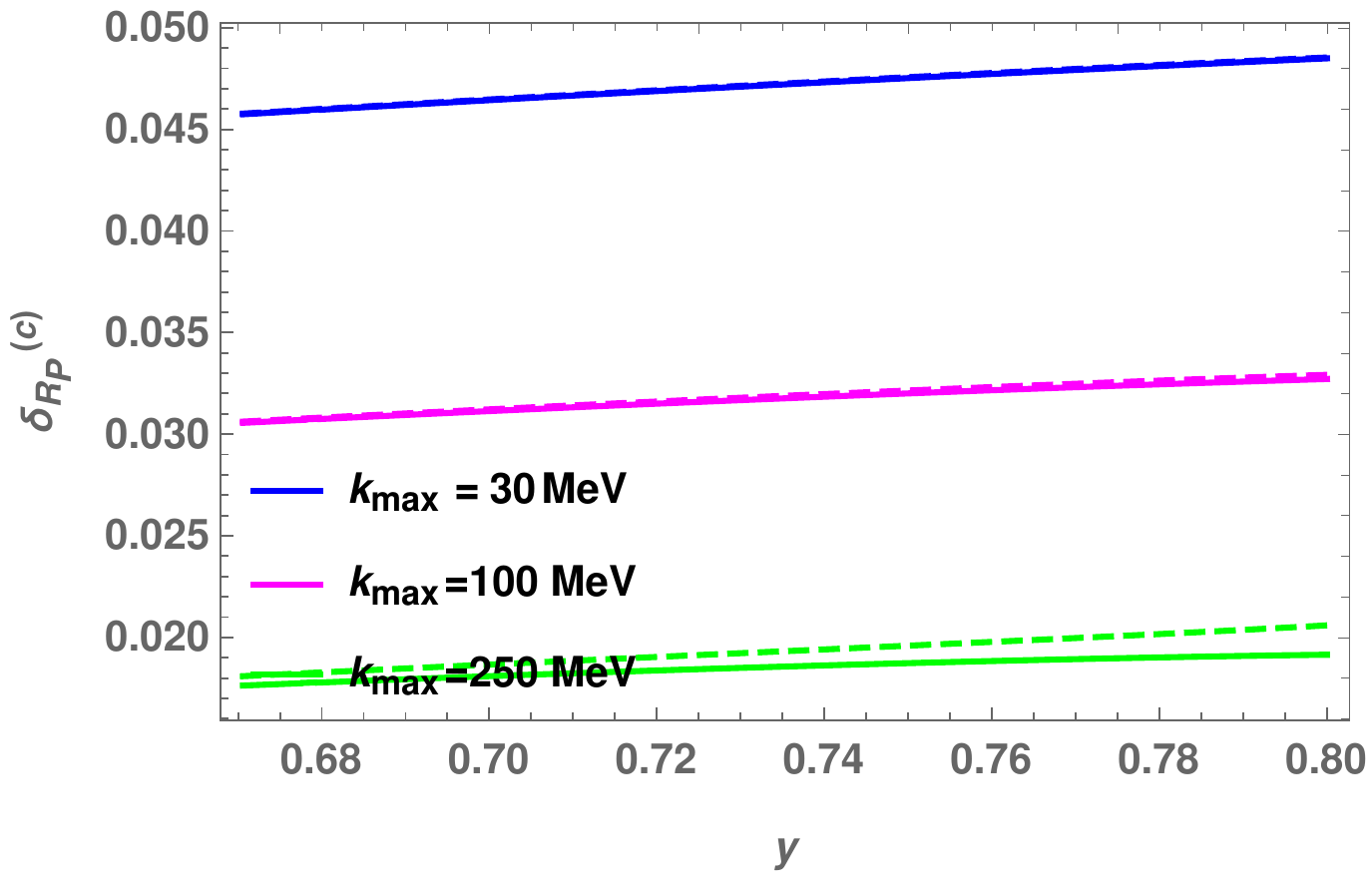}
	 		\caption{}
	 	\end{subfigure}
	 	\caption{Radiative corrections to $R_{P}$ ($P=D$ (dashed) and $\pi$(solid)) for different thresholds on photon energy, $k_{max}$, for (a) $B^0 \rightarrow P^+(=D^+,\pi^+) \ell^- \nu_\ell$ and (b) $B^- \rightarrow P^0(=D^0,\pi^0) \ell^- \nu_\ell$. }
	 	\label{fig7}
	 \end{figure}

	Next, we check the dependence or sensitivity of $R_V$ on the choice of form factors adopted for $B\to \pi$ and $B\to D$
	transitions. For this purpose, we chose two sets of form factors for both $B\to \pi$ and $B\to D$ process: 
	$(I)$ the form factors considered for the present analysis (see, Appendix- \ref{appb} for details),
	and $(II)$ the form factors obtained from the lattice analysis \cite{Aoki:2021kgd}. The sensitivity of $R_V$ on the choice of 
	form factors is tabulated in Table (\ref{tabel3}). For all these determinations,
	we limit ourselves to the large $q^2$ region such that the reliability of the chosen form factors is not questioned and a meaningful
	comparison between the different combinations formed is feasible. It can be seen from the table that there is very little impact on
	the choice of form factors. \\
	As $R_V$ turns out to be rather robust against soft photon corrections as well as choice of form factors, 
	it can thus be considered as a promising observable.\\
For completeness, we also consider the effect of soft photons on the flavour universality ratios, $R_{P(=D,\pi)}$ (shown in Fig. (\ref{fig7})). It is found that the soft photons lead to a shift of $\sim 2\%$ for $k_{max}= 250$ MeV for both the ratios. 
	 \subsection{Phenomenological application of $R_V$}
   Having demonstrated the insensitivity of $R_V $ to soft photon QED corrections as well as choice of form factors, we explore the 
   potential of $R_V$ in probing physics beyond the SM. To this end, and to keep the discussion simple but bring out the essence, 
   we consider new physics (NP) in the form of right handed currents in the quark sector given by the following effective Hamiltonian 
    \begin{eqnarray}
    H_{\text{NP}}=\frac{4G_F}{\sqrt{2}}V_{qb} c_R^q (\bar{\ell}\gamma_{\mu} P_L \nu)\left( \bar{q}\gamma_{\mu} P_R b \right),
    \label{eqnhnp}
    \end{eqnarray}
    where $q=u,c$ while $c_R^q$ are the Wilson coefficients. The contribution of the Hamiltonian given in eq. (\ref{eqnhnp}) to the differential decay width for the exclusive $B\to P \ell \bar{\nu}_{\ell}$ is given by 
    \begin{eqnarray}
    	\frac{d^{2}\Gamma_{B\to P \ell \bar{\nu}_{\ell}}}{dy}=\frac{d^{2}\Gamma_{B\to P \ell \bar{\nu}_{\ell}}}{dy}\Big|_{\text{SM}} |1+c_R^q|^2.
    \end{eqnarray}
    and for the inclusive case with $m_u/m_b \to 0$, it is
    \begin{eqnarray}
    	\frac{d^{2}\Gamma_{B\to X_q \ell \bar{\nu}_{\ell}}}{dy}=|1+c_R^q|^2\frac{d^{2}\Gamma_{B\to X_q \ell \bar{\nu}_{\ell}}}{dy}\Big|_{\text{SM}} +c_R^q\frac{d^{2}\Gamma_{B\to X_q \ell \bar{\nu}_{\ell}}}{dy}\Big|_{\text{LR}}.
    \end{eqnarray}
    The explicit expressions of $\frac{d^{2}\Gamma_{B\to X_q \ell \bar{\nu}_{\ell}}}{dy}\Big|_{\text{SM,LR}}$ can be found in 
    \cite{Dassinger:2008as,Kamali:2018bdp}.\\
    In the presence of this NP, we can extract $V_{ub}$ and $V_{cb}$ from different $b\to u$ and $b\to c$ modes (inclusive and exclusive), and these are tabulated in Table  (\ref{tabel1}). Here $V_{qb}^{\text{SM}}$ is the corresponding CKM element extracted if there were no NP.
    	\begin{table}[h]
    	\centering
    	\renewcommand{\arraystretch}{2}
    \begin{tabular}{ |c| c|c| }
    \hline
    {}&{ Modes } & {$V_{qb}^{NP} $} \\ \hline
\multirow{4}{*}{Exclusive Decays}   &{$B\to D \ell \nu_{\ell}$}& {$V_{cb}^{NP}=\frac{V_{cb}^{(\text{SM})}}{1+c_R^c }$}   \\

    &{$B\to D^* \ell \nu_{\ell}$}& {$V_{cb}^{NP}=\frac{V_{cb}^{(\text{SM})}}{1-c_R^c }$}   \\ 

    &{$B\to \pi \ell \nu_{\ell}$}& {$V_{ub}^{NP}=\frac{V_{ub}^{(\text{SM})}}{1+c_R^u }$}   \\ 
    
    &{$B\to \rho \ell \nu_{\ell}$}& {$V_{ub}^{NP}=\frac{V_{ub}^{(\text{SM})}}{1-c_R^u }$}   \\ \hline
    \multirow{2}{*}{Inclusive Decay}&{$ B\to X_c \ell \nu_{\ell}$}& {$V_{cb}=\frac{V_{cb}(\text{SM})}{1-0.34c_R^c }$}   \\ 
    
    &{$B\to X_u \ell \nu_{\ell}$}& {$V_{ub}=V_{ub}^{(\text{SM})} \hspace{0.3cm}(\text{for} \hspace{0.2cm}m_u\sim 0)$}   \\ \hline
    \end{tabular}
	    \caption{$V_{qb}^{NP}$ for various exclusive and inclusive $B$ decay modes }
	    \label{tabel1}
	\end{table}
Such NP contributions have impact on the observable $R_V$ as well. The ratio of $R_V^{NP}$ to $R_V^{\text{SM}}$ using various combinations of different channels are tabulated in Table \ref{table2}.
    	\begin{table}[h]
    	\centering
    	\renewcommand{\arraystretch}{2}
    \begin{tabular}{ |c| c| c|c|c|c|}
    \hline
    {  } & { $\frac{B\to X_u}{B\to X_c}$} & {$\frac{B\to \pi}{B\to D}$} & {$\frac{B\to \pi}{B\to D^*}$} & {$\frac{B\to \rho}{B\to D}$} & 
    {$\frac{B\to \rho}{B\to D^*}$}\\
    \hline
    {$\left(\frac{|V_{ub}|}{|V_{cb}|}\right)^{NP}/\left(\frac{|V_{ub}|}{|V_{cb}|}\right)_{\text{SM}}$}& {$1-0.34 c_R^c$} & {$1+c_R^c-c_R^u$} & {$1-c_R^c-c_R^u$} & {$1+c_R^c+c_R^u$} & {$1-c_R^c+c_R^u$}    \\
    \hline
    \end{tabular}
	    \caption{Ratio of $R_V$ in the NP to $R_V$ in the SM for inclusive $B\to X_u/ X_c$ modes and 
	    four different combination of exclusive $B\to \pi/D/\rho/D^*$ modes }
	    \label{table2}
	\end{table}
	As the ratio $R_V$ formed using the inclusive and exclusive determinations are equal (as discussed in Section- \ref{sec1}), this equality can be used to find
	constraints on new physics. On equating the ratio from the inclusive modes (i.e. first column) to the ratio from the exclusive modes, we get the constrains on  $c_R^u$ to be $c_R^u \in [-1.34,1.34]c_R^c$. 
	This shows the actual probing power of $R_V$: the up-quark right handed strength gets tightly correlated to the charm quark new
	physics coupling. While this example was a simple one, it is not difficult to convince oneself that $R_V$ holds similar power 
	in case of other new physics modifications as well.  \\
	Usually, in a model independent approach to $V_{cb}$ and $V_{ub}$ puzzles, the new physics couplings
	 in the two modes are treated independently. In specific models however, it may not be so. The equality of $R_V|_{incl}$
	 and $R_V|_{excl}$ leads to (simple) relations between the two couplings even when employing a model independent approach.\\
As a further phenomenological application,  we attempt to find the constraint on  
$\mathcal{BR}(B_c\to \tau \nu_{\tau}) $ using $ \mathcal{BR}(B\to \tau \nu_{\tau})$. The Branching ratio of $B(B_c) \to \tau \nu_{\tau}$
for the considered NP model is given by 
\begin{eqnarray}
\mathcal{BR}(B(B_c) \to \tau \nu_{\tau})&=&(1-2c_R^{u(c)}) \mathcal{BR}(B(B_c) \to \tau \nu_{\tau})|_{\text{SM}}
\end{eqnarray}
where,
\begin{eqnarray}
\mathcal{BR}(B(B_c)\to \tau \nu_{\tau})|_{\text{SM}}&=&\tau_{B(B_{c})}\frac{G_F^2 m_{B(B_c)} m_\tau^2}{8\pi}\left(1-\frac{m_{\tau}^2}{m_{B(B_c)}^2}\right) f_{B(B_c)}^2 |V_{u(c)b}|^2 
\end{eqnarray}
with $f_{B(B_c)}=185(434 )\text{MeV}$ being the decay constant for $B(B_c)$ meson. 
$\mathcal{BR}(B \to \tau \nu_{\tau})|_{\text{exp}}=1.09\times 10^{-4}$ \cite{Zyla:2020zbs} can be used to get the value of $c_R^u$. 
Using the obtained value of $c_R^u$, the estimated branching ratio for $B_c \to \tau \nu_{\tau}$ is found to be $[1.9-2.4]\%$ which is well
below the bound for $\mathcal{BR}(B_c \to \tau \nu_{\tau})\leq30\%$ as given in \cite{Alonso:2016oyd,Fleischer:2021yjo}.
If instead, the $\mathcal{BR}(B_c \to \tau \nu_{\tau})$ branching ratio had been greater than the bound, such new physics couplings
would have been disfavoured. This would not only have strained the resolution of $V_{cb}$ puzzle but also $V_{ub}$ puzzle since NP in up type quarks gets intimately related
to those in charm sector.\\
While we have explicitly checked insensitivity of $R_V$ to soft photon corrections and form factors, similar checks for the inclusive case
are beyond the scope of this work and are left for future study. For the present purpose, we assume that $R_V|_{incl}$ also presents similar
independence to such effects.

 \section{Conclusions and discussions}
	 \label{sec5}
	 $|V_{cb}|$ and $|V_{ub}|$ have consistently shown discrepancies when determined from exclusive and inclusive determinations. Due to
	 hadronic uncertainties involved, such discrepancies can't be confidently ascribed to physics beyond the SM. What about other
	 potential sources of uncertainty?  
	 We begin our discussion by investigating the impact of soft photon corrections to the determination of $|V_{cb}|$ and $|V_{ub}|$ 
	 considering the $B \rightarrow P \ell \nu_\ell$ decay processes, with $P=D,\,\pi$. We find that these elements get significant 
	 shift (roughly 3-4 $\%$) due to these QED corrections. To be more explicit (say, at $k_{max}=100$ MeV), we observe a correction of $\sim 2.2\%$ ($\sim 3.5\%$) for the charged $B$ to $D$ ($\pi$) mode whereas  for the case of neutral $B$, both $D$ and $\pi$ modes receive $\sim 1.7\%$ correction. These results are found to be in good agreement with other studies of the similar
	 nature.
	 In order to calculate the decay width, 
	 we consider the radiative corrections from inside as well as outside the Dalitz region for both the neutral and charged decay modes.
	 The corrections are found to be sensitive to the maximum energy, $k_{max}$ of the photon considered and very little sensitive to the 
	 angle between the lepton and the photon. Therefore, they are almost free from collinear divergences. The total QED correction
	 to the muon and tau channels are found to be $\sim -3.4\%$ and $\sim -1\%$, respectively for $k_{max} = 100$ MeV for neutral $B$ mode. While these corrections
	 are of the order of few percent corrections, such effects are still worrisome when aiming at less than a percent precision, and need to be
	 properly accounted for.\\
	 In order to find an observable free from the QED and the hadronic uncertainties, we suggest the use of the ratio of these CKM element 
	 (i.e. $R_V = \frac{|V_{ub}|}{|V_{cb}|}$) as a clean probe of the SM. We found that this ratio gets negligible correction due 
	 to the soft photon QED effects. Also, we have explicitly checked the impact of the choice of the form factors by considering different parametrizations or
	 choices. We found that this ratio, when evaluated in a judiciously chosen $q^2$ range, is affected very mildly 
	 by the choice of the form factors. Another intriguing observation is the excellent agreement between the $R_V$ values determined
	 from exclusive and inclusive determinations. Thus, while the individual CKM elements show puzzling behaviour and are sensitive
	 to QED as well as hadronic effects, the ratio, $R_V$, practically turns out to be insensitive to any such effects. These observations
	 together motivate $R_V$ to be a very useful observable in our quest for physics beyond the SM. Not only is $R_V$ a clean observable,
	 but the near perfect agreement of inclusive and exclusive determinations of $R_V$ allows one to equate the theoretically computed
	 expressions for the inclusive and exclusive cases. Equating these lead to simple relations between the new physics in
	 the $b\to u$ and $b\to c$ semi-leptonic modes. In the usual treatments of $|V_{cb}|$ and $|V_{ub}|$ puzzles, the new physics couplings
	 in the two modes are treated independently. However, in concrete models, the two are related in some way. The equality of $R_V|_{incl}$
	 and $R_V|_{excl}$ immediately relates the two type of coulings even in a model independent approach. These relations can then
	 be checked in concrete models to identify specific models which can address these puzzles.
	 We are thus encouraged to propose using $R_V$ in our quest for probing the SM itself and searching for new physics,
	 both experimentally and phenomenologically. 
	 
\appendix
\section{Kinematics and notations}
		\label{appa}
		\subsection{Three body kinematics }
		Kinematics for the three body decay $B\to P \ell \nu_{\ell}$
		 can be given in terms of three Lorentz invariant kinematic variables $x$, $y$ and $z$ or mandelstam variables $s$, $t$ and $u$. The kinematic variables are 
		\begin{eqnarray}
		x=\frac{Q^{2}}{m_{B}^{2}}, \hspace{0.3cm} y=\frac{2 p_{B}.p_{l}}{m_{B}^{2}}, \hspace{0.3cm}  z=\frac{2 p_{B}.p_{P}}{m_{B}^{2}}
		\end{eqnarray}
		where $Q^{2}=(p_{B}-p_{D}-p_{\ell})^{2}$. Note that $Q^{2}$ is zero in this process since it is the mass of the neutrino but it plays an important role when we discuss the real photon emission case. There it is defined as missing mass $(Q^{2}=(p_{\nu}+k)^{2})$ and yields non-zero value. The non-radiative decay width for $B\rightarrow P\ell \nu_\ell$ is given by
		{\small
		\begin{eqnarray}
		\Gamma_{0}
		&=& \frac{m_{B}}{256 \pi^{3}}\int dz \int dy \left|\mathcal{M}\right|_{B\to P \ell \nu_{\ell}}^{2}.
		\end{eqnarray}}
		One can see that the final result is independent of $x$ and therefore require only two independent Lorentz invariant kinematic variables $y$ and $z$. The kinematic boundaries for the variables $x$, $y$ and $z$ are: $x_{-}\leq x \leq x_{+} , \hspace{0.3cm} z_{-}\leq z \leq z_{+}  , \hspace{0.3cm} y_{-}\leq y \leq y_{+}$
		{\small
		\begin{eqnarray}
		\text{where,}\hspace{1cm}x_{\pm}&=& 1-y-z+\frac{m_{P}^{2}}{m_{B}^{2}}+\frac{m_{\ell}^{2}}{m_{B}^{2}}+\frac{yz}{2}\pm \frac{1}{2} \sqrt{y^{2}-4\frac{m_{l}^{2}}{m_{B}^{2}}}\sqrt{z^{2}-4\frac{m_{P}^{2}}{m_{B}^{2}}},\nonumber\\
		z_{\pm}&=& \frac{(2-y)(1+\frac{m_{P}^{2}}{m_{B}^{2}}+\frac{m_{\ell}^{2}}{m_{B}^{2}}-y)}{2(1+\frac{m_{\ell}^{2}}{m_{B}^{2}}-y)}\pm \frac{\sqrt{y^{2}-4\frac{m_{l}^{2}}{m_{B}^{2}}}(1-\frac{m_{P}^{2}}{m_{B}^{2}}+\frac{m_{\ell}^{2}}{m_{B}^{2}}-y)}{2(1+\frac{m_{\ell}^{2}}{m_{B}^{2}}-y)},\nonumber\\
		y_{-}&=& 2\sqrt{r_{\ell}}, \text{    and} \hspace{1.5cm}  y_{+}=1-\frac{m_{P}^{2}}{m_{B}^{2}}+\frac{m_{\ell}^{2}}{m_{B}^{2}}.\nonumber
		\end{eqnarray}
		}
		\subsection{Four body kinematics}
		The decay width for the process $B\to P \ell \nu_{\ell} \gamma$
		is given in terms of ten Lorentz invariant kinematic variables out of which five variables are independent and they are choosen as $x$, $y$, $z$, $p_{\nu}$ and $k$. The four body decay region is divided into two regions: $\mathcal{D}_{3}$ and $\mathcal{D}_{4-3}$. The decay width in these two regions is given by 
		\begin{eqnarray}
		\Gamma_{\mathcal{D}_3}|_{B\to P \ell \nu_{\ell} \gamma} &=&\frac{m_{B}^{3}}{512 \pi^{4}}\int_{\mathcal{D}_{3}} dy dz \int_{\frac{m_{\gamma}^{2}}{m_{B}^{2}}}^{x_{+}} dx  \int \frac{d^{3}p_{\nu}}{(2\pi)^{3} 2E_{\nu}}  \int \frac{d^{3}k}{(2\pi)^{3} 2E_{k}}(2\pi)^{4} \delta^{4}\left(Q- p_{\nu}\right.\nonumber\\&-&\left.k\right) \left|\mathcal{M}\right|_{B\to P \ell \nu_{\ell} \gamma}^{2}, \text{ and}\\
		\Gamma_{ \mathcal{D}_{4-3}}|_{B\to P \ell \nu_{\ell} \gamma}&=&\frac{m_{B}^{3}}{512 \pi^{4}}\int_{\mathcal{D}_{4-3}} dy dz \int_{x_{-}}^{x_{+}} dx  \int \frac{d^{3}p_{\nu}}{(2\pi)^{3} 2E_{\nu}}  \int \frac{d^{3}k}{(2\pi)^{3} 2E_{k}}(2\pi)^{4} \delta^{4}\left(Q- p_{\nu}\right.\nonumber\\&-&\left.k\right) \left|\mathcal{M}\right|_{B\to P \ell \nu_{\ell} \gamma}^{2},
		\end{eqnarray}
		respectively. Here the kinematic boundaries for $y$ and $z$ in the region $\mathcal{D}_{4-3}$ are 
			\begin{eqnarray}
		z_{-}&=&2\sqrt{\frac{m_{P}^{2}}{m_{B}^{2}}},\hspace{1cm} z_{+}=\frac{(2-y)(1+\frac{m_{P}^{2}}{m_{B}^{2}}+\frac{m_{\ell}^{2}}{m_{B}^{2}}-y)}{2(1+\frac{m_{\ell}^{2}}{m_{B}^{2}}-y)}- \frac{\sqrt{y^{2}-4\frac{m_{l}^{2}}{m_{B}^{2}}}(1-\frac{m_{P}^{2}}{m_{B}^{2}}+\frac{m_{\ell}^{2}}{m_{B}^{2}}-y)}{2(1+\frac{m_{\ell}^{2}}{m_{B}^{2}}-y)},\nonumber\\
		y_{-}&=& 2\sqrt{\frac{m_{\ell}^{2}}{m_{B}^{2}}}, \text{ and} \hspace{1.5cm} y_{+}=1-\frac{m_{P}^{2}}{m_{B}^{2}}+\frac{\frac{m_{\ell}^{2}}{m_{B}^{2}}}{1-\sqrt{\frac{m_{P}^{2}}{m_{B}^{2}}}}.\nonumber 
		\end{eqnarray}
		\section{Form factors}
		\label{appb}
		The form factors involved in the $B\rightarrow D\ell \nu_\ell$ in the model independent parametrization are given by \cite{Bernlochner:2017jka}
		\begin{eqnarray}
	f_{+}^D(q^2)&=&\frac{1}{\sqrt{r}}\left[(1+r) h_{+}-(1-r)h_{-}\right],\nonumber\\
	f_{-}^D(q^2)&=&\frac{1}{\sqrt{r}}\left[(1+r) h_{-}-(1-r)h_{+}\right], \text{ and}\nonumber\\
	f_{0}^D(q^2)&=& f_{+}^D(q^2)+\frac{1+r^{2}-2 r w}{1-r^{2}}f_{-}^D(q^2).\nonumber
	\end{eqnarray}
	Here $r=\frac{m_{D}}{m_{B}}$, $w=\frac{p_{B}.p_{D}}{m_{B}m_{D}}$, 
		\begin{eqnarray}
	h_{+}&=&\xi \left[1+\frac{\alpha}{\pi}\left(C_{V_{1}}+\frac{1+w}{2} \left(C_{V_{2}}+C_{V_{3}}\right)\right)+\left(\epsilon_{c}-\epsilon_{b}\right)L_{1} \right],  \text{ and}\nonumber\\
	h_{-}&=&\xi \left[\frac{\alpha}{\pi}\frac{1+w}{2} \left(C_{V_{2}}-C_{V_{3}}\right)+\left(\epsilon_{c}-\epsilon_{b}\right)L_{4} \right]\nonumber
	\end{eqnarray}
	with $z=\frac{m_{c}}{m_{b}}$, $L_{1}=0.72(w-1)$, $L_{4}=0.24$, $\epsilon_{c}=0.1807$, $\epsilon_{b}=0.0522$, $\xi=\left(\frac{2}{1+w}\right)^{2}$,
	\begin{eqnarray}
	C_{V_{1}}&=&\frac{1}{6z(w-w_{z})}\left[2(w+1)\left((3w-1)z-z^{2}-1\right)r_{w}+(12z(w_{z}-w)-(z^{2}-1)\log z)+4z (w-w_{z})\Omega \right],\nonumber\\
	C_{V_{2}}&=&\frac{-1}{6z^{2}(w-w_{z})^{2}}\left[\left(4w^{2}+2w)z^{2}-(2w^{2}+5w-1)z-(w+1)z^{3}+2\right) r_{w}+z\left(2(z-1)(w_{z}-w)+ \right)\right], \nonumber\\
	C_{V_{3}}&=&\frac{1}{6z(w-w_{z})^{2}}\Big[\left((2w^{2}+5w-1)z^{2}-(4w^{2}+2w)z-2z^{3}+w+1\right)r_{w}+\Big((3-2w)z^{2}+(2-4w)z\nonumber\\&+& 1\Big)\log z + 2z(z-1)(w_{z}-w)\Big].\nonumber 
	\end{eqnarray}
	 with $r_{w}=\frac{\log(w_{+})}{\sqrt{w^{2}-1}}$, $w_{z}=\frac{1}{2}\left(z+\frac{1}{2}\right)$ and
	\begin{eqnarray}
	\Omega=\frac{w}{2\sqrt{w^{2}-1}}\left[2 Li_{2}(1-w_{-}z)-2Li_{2}(1-w_{+}z)+Li_{2}(1-w_{+}^{2})-Li_{2}(1-w_{-}^{2})\right]- w r_{w} \log z+1.\nonumber
	\end{eqnarray}
	Here, $w_{+}=w+\sqrt{w^{2}-1}$, $w_{-}=w-\sqrt{w^{2}-1}$.\\
	For $B\to \pi \ell \nu_{\ell}$, the form factors in z-expansion parametrization are given by,\cite{Khodjamirian:2011ub}:
		\begin{eqnarray}
		f_+^\pi(q^2)&=&\frac{f_+(0)^\pi}{1-\frac{q^2}{m_{B*}^2}}\left\lbrace 1+ \sum_{k=1}^{N-1} b_k \left(z(q^2,t_0)^k-z(0,t_0)^k-(-1)^{N-K}\frac{k}{N}\left[z(q^2,t_0)^N-z(0,t_0)^N\right] \right)\right\rbrace, \text{ and}\nonumber\\
		f_0^\pi(q^2)&=&f_0^\pi(0)\left\lbrace 1+ \sum_{k=1}^{N} b_k^0 \left(z(q^2,t_0)^k-z(0,t_0)^k\right)\right\rbrace.\nonumber
		\end{eqnarray}
	Here,	$z(q^2,t_0)=\dfrac{\sqrt{(m_B+m_\pi)^2-q^2}-\sqrt{(m_B+m_\pi)^2-t_0}}{\sqrt{(m_B+m_\pi)^2-q^2}+\sqrt{(m_B+m_\pi)^2-t_0}}$, $f_{0}^{\pi}(0)=f_{+}^{\pi}(0)=0.281$, $b_1=-1.62$, $b_1^0=-3.98$ and $	t_0=(m_B+m_\pi)^2-2\sqrt{m_B m_\pi}\sqrt{(m_B+m_\pi)^2-q^2}$.
		\section{ Photon inclusive: Computational details}
		\label{appc}
		Here, we list the coefficients , $C_{m,n}$ and the integrals, $I_{m,n}$ for $\{m,n\}\in \{-2,2\}$, encountered in determination of the differential decay width for the inclusive photon case.
			{\small
		 \begin{eqnarray}
		C_{1,1} &=& 2 x y m_{B}^{4}((-3 f_{-}^{2}+2 f_{-} f_{+}+f_{+}^{2}) m_{\ell}^{2}- 4 f_{+} m_{B}^{2}(f_{-} y + f_{+})),\nonumber\\
		C_{1,-1}&=& 16 f_{+} m_{B}^{2} (f_{-} - f_{+})(y+z), \hspace{2cm} C_{-2,2} = 64f_{+}^{2} m_{\ell}^{2},\nonumber\\
			C_{-1,1}&=&- 32 f_{+} \left(f_{+}m_{B}^{2}(x+2y+z-1)-m_{D}^{2}f_{+}-(f_{-}-2f_{+}) m_{\ell}^{2} \right),\nonumber\\
		C_{2,-1}&=& -16 f_{+} m_{B}^{4} (f_{-}+f_{+})\left(2x+y+z-2\right),\hspace{0.5cm}	C_{2,0} = 8xm_{B}^{4}(f_{-}+f_{+}) \left(f_{+}ym_{B}^{2}+(f_{-}-f_{+})m_{\ell}^{2}\right)\nonumber\\
		C_{-1,2} &=& -16 f_{+} m_{\ell}^{2}\left(m_{B}^{2}(f_{-}(x+z-1)+f_{+}(-x+2y+z-3))-(f_{-}-f_{+})(m_{D}^{2}-m_{\ell}^{2}) \right), \nonumber\\
			C_{1,0} &=& - 4 m_{B}^{2} \Big[-2f_{+}m_{B}^{2}(f_{-}(3xy+4x+4z-4)+f_{+} x(y+4)+ 2f_{+}(y+1)(y+z-2)) +8f_{-}f_{+} m_{D}^{2}  \nonumber\\
		&+&m_{\ell}^{2}\left( f_{-}^{2}(y+z-2)-2f_{-}f_{+}(y+z+2)+ f_{+}^{2}(y+z-2)\right)
	\Big],\nonumber\\
		C_{0,1}&=& 4\Big[m_{B}^{2}\left(m_{\ell}^{2}(f_{-}^{2}(x+z-1)-2f_{-} f_{+} (3x + 2y + z - 1)+ f_{+}^{2}(5x+z+3)\right)\nonumber\\&-&4 f_{+}m_{B}^{4}(f_{-} y (x+z-1) + f_{+} (x(y-1)+2y(z-2)-z+1))+  (f_{-}+f_{+})^{2}m_{\ell}^{2}(m_{\ell}^{2}-m_{D}^{2}) \Big],\nonumber\\
		C_{0,2} &=& 4 x m_{B}^{2} m_{\ell}^{2} \Big[2f_{+}m_{B}^{2}(f_{-} y - f_{+}(y-2)) +(f_{-}-f_{+})^{2} m_{\ell}^{2}\Big],\nonumber\\
		C_{0,0} &=& -16f_{+} \Big[ m_{B}^{2}(f_{-}(x+2y+z-1) -f_{+}(x+4y+3z-1)) - (f_{-}-f_{+}) (-m_{D}^{2}-m_{\ell}^{2})\Big],	\nonumber     
		\end{eqnarray}
		{\small
		\begin{eqnarray}
		I_{0,0} &=& \frac{1}{4},\hspace{1.5cm} I_{1,1}= \frac{1}{4} \frac{2}{Q^{2}(p_{B}.p_{\ell}) \beta_{B \ell}}\log\left(\frac{1+\beta_{B \ell}}{1-\beta_{B \ell}}\right),\nonumber\\
		I_{2,0}&=& \frac{1}{m_{B}^{2}Q^{2}},\hspace{1cm}I_{1,0}= \frac{1}{4 (p_{B}.Q)\beta_{B Q}}\log\left(\frac{1+\beta_{B Q}}{1-\beta_{B Q}}\right) ,\nonumber\\
		I_{1,-1} &=& \frac{1}{4}\left(\frac{p_{B}p_{\ell}:Q}{(p_{B}.Q)^{2}\beta_{B Q}^{2}}+\frac{Q^{2} (p_{\ell} Q: p_{B})}{2(p_{B}.Q)^{3}\beta_{B Q}^{2}}\log\left(\frac{1+\beta_{B Q}}{1-\beta_{B Q}}\right)\right),\nonumber\\
		I_{2,-1}&=& \frac{1}{4}\Big(\frac{2(p_{\ell}Q:p_{B})}{m_{B}^{2}(p_{B}.Q)^{2}\beta_{B Q}^{2}}+\frac{(p_{B}p_{\ell}:Q)}{(p_{B}.Q)^{3} \beta_{B Q}^{3}}\log\left(\frac{1+\beta_{B Q}}{1-\beta_{B Q}}\right)\Big), \text{ and}\nonumber\\
		I_{-2,2}&=&\frac{1}{4}\Big[\frac{Q^{2}(p_{B}Q:p_{\ell})^{2}}{m_{\ell}^{2}(p_{\ell}.Q)^{4}\beta_{\ell Q}^{4}}+\frac{Q^{2}(p_{B}Q:p_{\ell})(p_{B}p_{\ell}:Q)}{(p_{\ell}.Q)^{5}\beta_{\ell Q}^{5}}\log\left(\frac{1+\beta_{\ell Q}}{1-\beta_{\ell Q}}\right)+\frac{(p_{B}p_{\ell}:Q)^{2}}{(p_{\ell}.Q)^{4}\beta_{\ell Q}^{4}}\nonumber\\
		&-&\frac{(p_{B}.Q)^{2}(p_{\ell}.Q)^{2}\beta_{B Q}^{2}\beta_{\ell Q}^{2}-(p_{B}p_{\ell}:Q)^{2}}{2(p_{\ell}.Q)^{4}\beta_{\ell Q}^{4}}\left(2-\frac{1}{\beta_{\ell Q}}\log\left(\frac{1+\beta_{\ell Q}}{1-\beta_{\ell Q}}\right)\right)\Big]. \nonumber
		\end{eqnarray}}
		Here, $\beta_{i j}=\sqrt{1-\frac{m_{i}^{2}m_{j}^{2}}{(p_{i}.p_{j})^{2}}}$,  
		$p_{i}p_{j}:p_{k}=(p_{i}.p_{k})(p_{j}.p_{k})-p_{k}^{2}(p_{i}.p_{j})$ and $I_{m,n}(p_{i},p_{j})= I_{n,m}(p_{j},p_{i})$. The integrals $I_{m,n}$ are found to be consistent with \cite{Ginsberg:1967gvl}.
    	}
    	
		\section{Integrals for real emission and virtual corrections}
		\label{appd}
		Here we list various integrals involved in the real photon emission for both photon inclusive and photon exclusive scenarios.
		 \subsection{Photon Inclusive case:}
		\begin{eqnarray}
		\int_{0}^{x_{+}} dx I_{1,1}&=& \frac{1}{4m_B^2} \int_{-1}^{1} dt \frac{1}{p_{t}^{2}}\log\left(\frac{x_{+}^2 p_t^2}{m_{\gamma}^2 E_t^2}\right)+\text{NIR},\nonumber\\
		\int_{0}^{x_{+}} dx I_{2,0}&=& \frac{1}{2m_{B}^{4}}\log\left(\frac{x_{+}^2}{m_{\gamma}^{2}}\right)+ \text{NIR}, \hspace{0.5cm}\text{and}\nonumber\\
		\int_{0}^{x_{+}} dx I_{0,2}&=& \frac{1}{2m_B^2 m_{\ell}^{2}}\log\left(\frac{x_{+}^2 }{m_{\gamma}^{2}}\right)+\text{NIR},\nonumber
		\end{eqnarray}
		where $\text{NIR}$ is contribution from finite non-IR terms. 
		\subsection{Exclusive photon case:}
		We have
		\begin{eqnarray}
		\tilde{B}&=& \frac{-1}{8\pi^{2}}\int_{0}^{k_{max}}\frac{d^{3}k}{(k^{2}+m_{\gamma}^{2})^{1/2}}\left[\frac{m_{i}^{2}}{(k.p_{i})^{2}}+\frac{m_{j}^{2}}{(k.p_{j})^{2}}-\frac{2p_{i}.p_{j}}{(k.p_{i})(k.p_{j})}\right].  
		\end{eqnarray}
		
		The various integrals involved are,
		\begin{equation}
		\int_{0}^{k_{max}}\frac{d^{3}k}{(k^{2}+m_{\gamma}^{2})^{1/2}}\frac{1}{(k.p_{i})^{2}}=\frac{2\pi}{m_{i}^{2}}\ln\left( \frac{k_{max}^{2}m_{i}^{2}}{E_{i}^{2}m_{\gamma}^{2}}\right), 
		\end{equation}
		\begin{eqnarray}
		\int_{0}^{k_{max}}\frac{d^{3}k}{(\vec{k}^{2}+m_{\gamma}^{2})^{1/2}}\frac{1}{(k.p_{i})(k.p_{j})}&=& \frac{2\pi}{2}\int_{-1}^{1}\frac{dt}{p_{t}^{2}}\ln\left( \frac{k_{max}^{2}p_{t}^{2}}{E_{t}^{2}m_{\gamma}^{2}}\right) + \text{finite term} 
		\end{eqnarray}
		
		\section{Useful functions involved in virtual photon corrections}
		\label{appe}
		The scalar two point and three point Passarino-Veltman functions and their derivatives are (with $m_{\gamma}$ and $\Lambda$ as IR and UV regulators):
		\begin{eqnarray}
		B_{0}(m_{a}^{2},0, m_{a}^{2})&=& 2-\ln\left(\frac{m_{a}^{2}}{\Lambda^{2}}\right)\label{3}, \hspace{0.5cm} \text{and}\\
		B_{0}(q^{2},m_{a}^{2}, m_{b}^{2}) &=& -\int_{0}^{1}du \ln\frac{-u(1-u)q^{2} + u m_{b}^{2}+(1-u)m_{a}^{2}}{\Lambda^{2}}\\
			B'_{0}(m_{i}^{2},m_{\gamma}^{2}, m_{i}^{2})&=& \frac{-1}{2m_{i}^{2}}\left(2+\ln\left(\frac{m_{\gamma}^{2}}{m_{i}^{2}}\right)\right)
		\end{eqnarray} 
		\begin{eqnarray}
		C_{0}(m_{B}^{2}, m_{\ell}^{2}, q^{2}, m_{B}^{2}, m_{\gamma}^{2}, m_{\ell}^{2}) &=&\frac{-1}{4} \int_{-1}^{1} dt \frac{1}{p_{t}^{2}}\ln\left(\frac{m_{\gamma}^{2}}{p_{t}^{2}}\right),\\
		C_{1}(m_{B}^{2}, m_{\ell}^{2}, q^{2}, m_{B}^{2}, 0, m_{\ell}^{2}) &=&\frac{1}{2m_{\ell}^2\beta^2}\Bigg[p_{\ell}^2\Big(B_{0}[m_{\ell}^2,0,m_{\ell}^2]-B_0[q^2,m_{\ell}^2,m_B^2]\Big)\nonumber\\&-& p_B.p_{\ell}\Big(B_{0}[m_B^2,0,m_B^2]- B_0[q^2,m_{\ell}^2,m_B^2]\Big)\Big], \hspace{0.5cm} \text{and}\\
		 C_{2}(m_{B}^{2}, m_{\ell}^{2}, q^{2}, m_{B}^{2}, 0, m_{\ell}^{2}) &=&\frac{1}{2m_{B}^2\beta^2}\Bigg[-p_B.p_{\ell}\Big(B_{0}[m_{\ell}^2,0,m_{\ell}^2]-B_0[q^2,m_{\ell}^2,m_B^2]\Big)\nonumber\\&-&p_B^2\Big(B_{0}[m_B^2,0,m_B^2]- B_0[q^2,m_{\ell}^2,m_B^2]\Big)\Big],
		\end{eqnarray} 
		respectively. Here, $\beta=\frac{|\bold{p_{\ell}}|}{E_{\ell}}$ is the charged lepton velocity in the rest frame of decaying particle. 

\bibliographystyle{ieeetr}
\bibliography{vub_vcb}

\end{document}